\documentclass[twocolumn]{article}
\usepackage{amsmath}
\usepackage{amsthm}
\usepackage{lodin}
\usepackage{multirow}
\usepackage{array}
\usepackage{graphicx}
\usepackage[nomargin]{fixme}
\usepackage{stmaryrd}
\usepackage{mathpartir}
\usepackage{hyperref}
\usepackage{mathtools}
\usepackage[english]{babel}
\usepackage{rotating}
\usepackage{lipsum}
\usepackage{adjustbox}
\usepackage{braket}
\usepackage{fontawesome}
\usepackage{algorithm2e}
\usepackage{backnaur}
\usepackage{mdframed}
\usepackage{subfig}
\usepackage{tikz}
\usepackage{multicol}
\usepackage{booktabs}
\usepackage{paralist}
\usepackage[numbers,sort]{natbib}
\usepackage{xspace}
\usepackage{float}
\usepackage{authblk}
\usepackage{url}

\usetikzlibrary{arrows,automata,shapes,patterns,matrix,shadows,fit,positioning,chains}
\usetikzlibrary{positioning, shapes,patterns,shadows.blur, arrows,decorations,arrows,automata,shadows,patterns,chains,graphs,calc,intersections,matrix,fit,shapes,chains,decorations.pathreplacing,chains,fit,shapes}

\lstdefinestyle{terminal}
{
    numbers=none,
    basicstyle=\scriptsize\color{white}\ttfamily,
    keywordstyle=\scriptsize\color{white}\ttfamily,
    commentstyle=\scriptsize\color{white}\ttfamily,
    identifierstyle=\scriptsize\color{white}\ttfamily,
    stringstyle=\scriptsize\color{white}\ttfamily,
    numberstyle=\scriptsize\color{white}\ttfamily,
    frame=none
  }

\newenvironment{cmd}{\begin{mdframed}[backgroundcolor=gray]}{\end{mdframed}}

\lstnewenvironment{term}
    {\lstset{style=terminal}}  
    {}

\lstnewenvironment{cprgm}
    {\lstset{style=customc}}  
    {}

\newfloat{lstllvm}{tbp}{lop}
\floatname{lstllvm}{\llvm-Listing}

\newfloat{lstterm}{tbp}{lop}
\floatname{lstterm}{\lodin-Output}

\title{Automatic Verification of LLVM Code}

\author[1]{Axel~Legay}
\author[2]{Dirk~Nowotka}
\author[3]{Danny~Bøgsted~Poulsen}
\affil[1]{UCLouvain, Belgium} 
\affil[2]{Kiel University, Germany}
\affil[3]{Aalborg University, Denmark}

\newcommand{\bitvector}[1]{\ensuremath{\mathbb{B}^{#1}}}

\newcommand{\typej}{\ensuremath{\,:\,}}

\newcommand{\domainname}[1][]{\ensuremath{\mathcal{A}_{#1}}}
\newcommand{\domainexplicit}{\ensuremath{\mathcal{E}}}
\newcommand{\domainsymbolic}{\ensuremath{\mathcal{S}}}
\newcommand{\domainstates}[1][\domainname]{\ensuremath{\mathcal{S}_{#1}}}
\newcommand{\domainstate}[1][]{\ensuremath{\mathsf{s}_{#1}}}    
\newcommand{\initdomainstate}[1][]{\ensuremath{{\domainstate[#1]}^{\mathtt{init}}}}
\newcommand{\domainfortypefname}[1]{\ensuremath{\mathtt{dom}_{#1}}}
\newcommand{\domainfortype}[2][\domainname]{\ensuremath{\domainfortypefname{#1}(#2)}}
\newcommand{\regvars}[1][\domainname]{\ensuremath{\mathcal{R}}}
\newcommand{\regvar}[1][]{\ensuremath{\mathsf{r_{#1}}}}

\newcommand{\val}{\ensuremath{v}}

\newcommand{\semadd}[2]{\ensuremath{+^{#1}_{#2}}}
\newcommand{\semsub}[2]{\ensuremath{-^{#1}_{#2}}}
\newcommand{\semmul}[2]{\ensuremath{\cdot^{#1}_{#2}}}
\newcommand{\semsdiv}[2]{\ensuremath{{/_s^{#1}}_{#2}}}
\newcommand{\semudiv}[2]{\ensuremath{{/_u^{#1}}_{#2}}}
\newcommand{\semsrem}[2]{\ensuremath{{\%_s^{#1}}_{#2}}}
\newcommand{\semurem}[2]{\ensuremath{{\%_u^{#1}}_{#2}}}
\newcommand{\semshl}[2]{\ensuremath{<<^{#1}_{#2}}}
\newcommand{\semlshr}[2]{\ensuremath{{>>_l^{#1}}_{#2}}}
\newcommand{\semashr}[2]{\ensuremath{{>>_a^{#1}}_{#2}}}
\newcommand{\semand}[2]{\ensuremath{\&^{#1}_{#2}}}
\newcommand{\semor}[2]{\ensuremath{|^{#1}_{#2}}}
\newcommand{\semxor}[2]{\ensuremath{\oplus^{#1}_{#2}}}

\newcommand{\semmakereg}[1]{\ensuremath{\mathbf{mReg}_{#1}}}

\newcommand{\semevalreg}[2][\ty{ty}]{\ensuremath{\mathbf{Eval}^{#1}_{#2}}}
\newcommand{\semalloc}[2]{\ensuremath{\mathbf{alloc}^{#1}_{#2}}}
\newcommand{\semfree}[1]{\ensuremath{\mathbf{free}_{#1}}}
\newcommand{\semload}[2]{\ensuremath{\mathbf{load}^{#1}_{#2}}}
\newcommand{\semstore}[2]{\ensuremath{\mathbf{store}^{#1}_{#2}}}
\newcommand{\semset}[2]{\ensuremath{\mathbf{Set^{#1}}_{#2}}}
\newcommand{\semnondet}[2]{\ensuremath{\mathbf{NonDet^{#1}_{#2}}}}
\newcommand{\semptradd}[1][\domainname]{\ensuremath{\mathbf{PtrAdd}_{#1}}}
\newcommand{\semeq}[2]{\ensuremath{==^{#1}_{#2}}}
\newcommand{\semne}[2]{\ensuremath{\neq^{#1}_{#2}}}
\newcommand{\semugt}[2]{\ensuremath{{>_u^{#1}}_{#2}}}

\newcommand{\semult}[2]{\ensuremath{{<_u^{#1}}_{#2}}}
\newcommand{\semule}[2]{\ensuremath{{\leq_u^{#1}}_{#2}}}
\newcommand{\semsgt}[2]{\ensuremath{{>_s^{#1}}_{#2}}}

\newcommand{\semslt}[2]{\ensuremath{{<_s^{#1}}_{#2}}}
\newcommand{\semsle}[2]{\ensuremath{{\leq_s^{#1}}_{#2}}}
\newcommand{\semtrue}[1]{\ensuremath{\mathtt{tt}_{#1}}}
\newcommand{\semfalse}[1]{\ensuremath{\mathtt{ff}_{#1}}}

\newcommand{\explicitdomain}{\ensuremath{\mathcal{E}}}

\newcommand{\denot}[1]{\ensuremath{\llbracket #1 \rrbracket}}
\newcommand{\apply}[2]{\ensuremath{\denot {#1}_{#2}}}
\newcommand{\instr}[1][]{\ensuremath{\op{Inst}_{#1}}}
\newcommand{\typerule}[3]{\ensuremath{\inferrule[#1]{#2}{#3}}}

\newcommand{\transrule}[4][]{\ensuremath{\inferrule[#1]{#3}{#4},\, {{\substack{#2}}}}}

\newcommand{\returnInsts}{\ensuremath{\mathtt{Rets}}}

\newcommand{\registermap}[1][]{{\ensuremath{\pi_{\mathsf{#1}}}}}
\newcommand{\frees}[1][]{\ensuremath{\mathtt{Free}_{#1}}}

\newcommand{\insttoop}{\ensuremath{\circ}}

\newcommand{\prev}[1][]{\ensuremath{\lllabel{prev}_{#1}}}
\newcommand{\cur}[1][]{\ensuremath{\lllabel{cur}_{#1}}}
\newcommand{\pc}[1][]{\ensuremath{\mathtt{pc}_{#1}}}

\newcommand{\instcmpop}[1][\ty{ty}]{\ensuremath{\gamma^{#1}}}

\newcommand{\stackk}[1][]{\ensuremath{\mathtt{SL}}}
\newcommand{\explisymb}{\ensuremath{\text{\domainexplicit}}}
\newcommand{\symbolicsymb}{\ensuremath{\text{\domainsymbolic}}}
\newcommand{\standardsymb}{\ensuremath{\explisymb}}
\newcommand{\tausymb}{\ensuremath{\text{\faBicycle}}}
\newcommand{\ntausymb}{\ensuremath{\text{\faBinoculars}}}

\newcommand{\systranss}[2][]{\ensuremath{\xrightarrow[#1]{#2}^{\domainname}}}
\newcommand{\systrans}[2][]{\ensuremath{\xrightarrow[#1]{#2}^{\standardsymb}}}
\newcommand{\systransexpli}[2][]{\ensuremath{\xrightarrow[#1]{#2}^{\explisymb}}}
\newcommand{\systranssymb}[2][]{\ensuremath{\xrightarrow[#1]{#2}^{\symbolicsymb}}}
\newcommand{\systranstau}[2][]{\ensuremath{\xrightarrow[#1]{#2}^{\tausymb}}}
\newcommand{\systransntau}[2][]{\ensuremath{\xrightarrow[#1]{#2}^{\ntausymb}}}
\newcommand{\actrecord}[1][]{\ensuremath{\kappa_{#1}}}
\newcommand{\stub}{\ensuremath{\mathtt{stub}}}
\newcommand{\emptylist}{\ensuremath{\epsilon}}

\newcommand{\boolv}{\ensuremath{\mathbb{B}}}

\newcommand{\unsignedBV}[1]{\ensuremath{\langle #1 \rangle}}
\newcommand{\signedBV}[1]{\ensuremath{\langle \cdot #1 \cdot \rangle}}
\newcommand{\unencodeBV}[1]{\ensuremath{\unsignedBV{#1}^{-1}}}
\newcommand{\signencodeBV}[1]{\ensuremath{\signedBV{#1}^{-1}}}

\newcommand{\bitadd}{\ensuremath{\, \mathtt{add}\, }}
\newcommand{\bitsub}{\ensuremath{\,\mathtt{sub}\,}}
\newcommand{\bitudiv}{\ensuremath{\,\mathtt{div_u}\,}}
\newcommand{\bitsdiv}{\ensuremath{\,\mathtt{div_s}\,}}
\newcommand{\bitmul}{\ensuremath{\,\mathtt{mul}\,}}
\newcommand{\biturem}{\ensuremath{\,\mathtt{rem_u}\,}}
\newcommand{\bitsrem}{\ensuremath{\,\mathtt{rem_s}\,}}

\newcommand{\bitand}{\ensuremath{\, \mathtt{and} \, }}
\newcommand{\bitor}{\ensuremath{\, \mathtt{or}\, }}
\newcommand{\bitneg}{\ensuremath{\, \mathtt{neg}\,  }}
\newcommand{\bitxor}{\ensuremath{\,  \mathtt{xor} \, }}
\newcommand{\bitlshl}{\ensuremath{\,  \mathtt{lshl} \, }}
\newcommand{\bitlshr}{\ensuremath{\,  \mathtt{lshr} \,  }}
\newcommand{\bitashr}{\ensuremath{\, \mathtt{ashr}} \, }
\newcommand{\bitconcat}{\,  \ensuremath{\circ} \, }
\newcommand{\zerovec}[1]{\ensuremath{\vec{0}^{#1}}}
\newcommand{\onevec}[1]{\ensuremath{\vec{1}^{#1}}}

\newcommand{\memblockmap}{\ensuremath{\mathtt{M}}}
\newcommand{\memblockmapinit}{\memblockmap_{\mathtt{init}}}
\newcommand{\memblock}{\ensuremath{MB}}
\newcommand{\memblockunused}{\ensuremath{\text{\faChainBroken}}}
\newcommand{\memused}{\ensuremath{\mathtt{Used}}}

\newcommand{\newmem}{\ensuremath{\mathtt{new}}}
\newcommand{\free}{\ensuremath{\mathtt{Memfree}}}
\newcommand{\rread}{\ensuremath{\mathtt{read}}}
\newcommand{\wwrite}{\ensuremath{\mathtt{write}}}

\newcommand{\dmin}{\ensuremath{\mathbf{min}}}

\newcommand{\state}[1][]{\ensuremath{s_{#1}}}

\newcommand{\prop}[1]{\ensuremath{\mathtt{#1}}}
\newcommand{\propfunc}[1]{\ensuremath{\mathcal{P}_{\prop{#1}}}}

\newcommand{\aritfunc}[1]{\ensuremath{\mathcal{A}_{#1}}}

\newcommand{\netstate}[1][]{\ensuremath{\mathsf{n}_{#1}}}
\newcommand{\netinitstate}[1][]{\ensuremath{\mathsf{n}^{0}_{#1}}}
\newcommand{\netstates}[1][]{\ensuremath{\mathsf{N}}}

\newcommand{\lang}[1]{\ensuremath{\Psi(#1)}}
\newcommand{\internalinst}[1]{\ensuremath{\mathtt{Internal(#1)}}}

\newcommand{\selectp}[1]{\ensuremath{\gamma_{#1}}}
\newcommand{\selectv}[1]{\ensuremath{\delta_{#1}}}
\newcommand{\prob}{\ensuremath{\mathbb{P}}}
\newcommand{\lts}[2]{\ensuremath{\mathcal{L}^{#1}_{#2}}}

\newcommand{\seahorn}{\textsc{SeaHorn}}
\newcommand{\cbmc}{\textsc{CBMC}}
\newcommand{\cpachecker}{\textsc{CPAChecker}}
\newcommand{\klee}{\textsc{Klee}}
\newcommand{\llbmc}{\textsc{LLBMC}}
\newcommand{\divine}{\textsc{Divine}}

\newcommand{\smtvar}[1][]{\ensuremath{v_{#1}}}
\newcommand{\smtite}[3]{\ensuremath{\mathtt{ite}(#1,#2,#3)}}
\newcommand{\smtvars}[1][]{\ensuremath{\mathcal{V}^{#1}}}
\newcommand{\smtexprs}[1][]{\ensuremath{\mathcal{W}^{#1}}}
\newcommand{\theory}{\ensuremath{\mathcal{T}}}

\newcommand{\arrselect}{\ensuremath{\mathtt{select}}}  
\newcommand{\arrstore}{\ensuremath{\mathtt{store}}}
\newcommand{\arrtype}[2]{\ensuremath{\{#1\}\rightarrow\{#2\}}}
\newcommand{\pathform}{\ensuremath{\psi}}
\newcommand{\bytesize}{\ensuremath{\mathtt{BSize}}}

\newcommand{\syntaxeq}[2]{\ensuremath{#1\stackrel{\mathtt{def}}{=} (#2)}}
\newcommand{\eg}{e.g.\xspace}
\newcommand{\ie}{i.e.\xspace}

\newcommand{\truee}{\ensuremath{\mathtt{tt}}}
\newcommand{\falsee}{\ensuremath{\mathtt{ff}}}
\newcommand{\incoming}{\ensuremath{\mathtt{In}}}

\newcommand{\libname}[1]{\ensuremath{\mathtt{#1}}}

\newcommand{\typeind}{\mathcal{T}}

\newcommand{\downloadurl}{\url{www.fillthis.later}}
\newcommand{\zthree}{\textsc{Z3}}
\newcommand{\boolector}{\textsc{Boolector}}

\newcommand{\binrule}{\ensuremath{\transrule[Binary]{
  s = 
    (\lodinfunction,\prev,\cur,\pc,\registermap,\frees)\\ \lodinfunction = (\funcname{N},\registers,\parameters,\blocklabels,\blocksanon,\blockmap,\ty{ret}), \\\syntaxeq{\instr}
    {\llbin{\%res}{inst}{ty}{\%inp1}{\%inp2}}\\
    \regvar[1] = \registermap(\reg{\%inp1}) \\
    \regvar[2] = \registermap(\reg{\%inp2}) \\
    \regvar[res] = \registermap(\reg{\%res})
  }
  {\instr = \blockmap(\cur)[\pc] \and
      \val \in \apply{\semevalreg{\domainname}(\domainstate,\regvar[1])) \,\insttoop(\op{inst})\,
      \semevalreg{\domainname}(\domainstate,\regvar[2]) }{\domainstate}
    }
    {\domainstate, \module\vdash (s,\stackk)
  \xrightarrow{\instr}
  ((\lodinfunction,\prev,\cur,\pc+1,\registermap,\frees),\stackk),\semset{\ty{ty}}{\domainname}
  (\domainstate,\regvar[res],\val)}})
}

\newcommand{\allocrule}{\ensuremath{\transrule[Alloc]{s = 
      (\lodinfunction,\prev,\cur,\pc,\registermap,\frees)\\ \lodinfunction = (\funcname{N},\registers,\parameters,\blocklabels,\blocksanon,\blockmap,\ty{ret}), \\\syntaxeq{\instr}{
    \llallocaone[ty]{\%res}} \\
    \regvar[res] = \registermap(\reg{\%res})
    }{\instr = \blockmap(\cur)[\pc] \and \semalloc{\ty{ty}}{\domainname} (\domainstate) = \domainstate',\val
      }{\domainstate, \module\vdash (s,\stackk)
  \xrightarrow{\instr}
  ((\lodinfunction,\prev,\cur,\pc+1,\registermap,\frees\cup\{m\}),\stackk),\semset{\ptr{ty}}{\domainname}
      (\domainstate',\regvar[res],\val)}
      }     
            }
        
\newcommand{\brunrule}{\ensuremath{
    \transrule[Branch Unconditional]{
    s = (\lodinfunction,\prev,\cur,\pc,\registermap,\frees)\\
    \syntaxeq{\instr}{\llbr{block}}\\
    \lodinfunction =
     (\funcname{N},\registers,\parameters,\blocklabels,\blocksanon,\blockmap,\ty{ret})
     }
     {\instr = \blockmap(\cur)[\pc]
     }
     {\domainstate, \module\vdash (s,\stackk)
     \xrightarrow{\instr}
     ((\lodinfunction,\cur,\lllabel{block},0,\registermap,\frees),\stackk),\domainstate}
}
}

\newcommand{\brcondtrue}{\ensuremath{
    \transrule[Branch Conditional True]{
  s = (\lodinfunction,\prev,\cur,\pc,\registermap,\frees)\\
  \syntaxeq{\instr}{\llbrcond{\%cond}{ttblock}{ffblock}} \\
  \lodinfunction =
    (\funcname{N},\registers,\parameters,\blocklabels,\blocksanon,\blockmap,\ty{ret})
}
{\instr = \blockmap(\cur)[\pc] \and
  \apply{\semevalreg{\domainname}(\domainstate,\registermap(\reg{\%cond})) \semne{\ty{ty}}{\domainname} \semfalse{\domainname}}{\domainstate} = \domainstate',\_,\top
}
{\domainstate, \module\vdash (s,\stackk)
  \xrightarrow{\instr}
  ((\lodinfunction,\cur,\lllabel{ttblock},0,\registermap,\frees),\stackk),\domainstate'}
    }
}

\newcommand{\brcondfalse}{\ensuremath{
    \transrule[Branch Conditional False]{
  s = (\lodinfunction,\prev,\cur,\pc,\registermap,\frees)\\
  \syntaxeq{\instr}{\llbrcond{\%cond}{ttblock}{ffblock}} \\
  \lodinfunction =
    (\funcname{N},\registers,\parameters,\blocklabels,\blocksanon,\blockmap,\ty{ret})
}
{\instr = \blockmap(\cur)[\pc] \and
  \apply{\semevalreg{\domainname}(\domainstate,\registermap(\reg{\%cond})) \semeq{\ty{ty}}{\domainname} \semfalse{\domainname}}{\domainstate} = \domainstate',\_,\top
}
{\domainstate, \module\vdash (s,\stackk)
  \xrightarrow{\instr}
  ((\lodinfunction,\cur,\lllabel{ffblock},0,\registermap,\frees),\stackk),\domainstate'}

    }
}

\newcommand{\cmpruleT}{\ensuremath{\transrule[Compare]{
    s = (\lodinfunction,\prev,\cur,\pc,\registermap,\frees)\\
    \lodinfunction =
    (\funcname{N},\registers,\parameters,\blocklabels,\blocksanon,\blockmap,\ty{ret}),\\
     \syntaxeq{\instr}{\llcmp[cond]{\ty{ty}}{\%res}{\%inp1}{\%inp2}} \\ 
     \instcmpop(\subop{cond}) = (op,\_) \\
     \regvar[1] = \registermap(\reg{\%inp1}) \\
     \regvar[2] = \registermap(\reg{\%inp2}) \\
}
{\instr = \blockmap(\cur)[\pc] \and
  \apply{\semevalreg{\domainname}(\domainstate,\regvar[1])\, op^{\ty{ty}}\, \semevalreg{\domainname}(\domainstate,\regvar[2])}{\domainstate} = \_,\val,\_
}
{\domainstate, \module\vdash (s,\stackk)
  \xrightarrow{\instr}
  (\lodinfunction,\prev,\cur,\pc+1,\registermap,\frees),\stackk),\semset{\ty{ty}}{\domainname}(\domainstate,\registermap(\reg{\%res}),\val)}}}

\newcommand{\callrule}{\ensuremath{
    \transrule[Call Function]{
    s = (\lodinfunction,\prev,\cur,\pc,\registermap[old] ,\frees)\\
    s' = (\lodinfunction,\prev,\cur,\pc+1,\registermap[old],\frees)\\
    \syntaxeq{\instr}{\op{call}\, \ty{ret}\, \reg{@func}\, (\ty{ty_0}\, \reg{\%\hat p_0} \dots
  \ty{ty_n}\, \reg{\%\hat p_n})}  \\ 
  \forall i, v_i = \semevalreg[\ty{ty_i}]{\domainname}(\domainstate,\registermap[old](\reg{\%\hat p_i})), \\
  \lodinfunction' = (\funcname{func},\registers,\parameters,\blocklabels,\blocksanon,\blockmap,\ty{ret})\in\module,\\
  \parameters = \{\reg{p_0},\dots,\reg{p_{n-1}}\},\\
  \registers = \{\reg{r_1},\dots,\reg{r_m}\},\\
  \registermap[0]' : \registers \rightarrow \regvars
}
{\instr = \blockmap(\cur)[\pc] \and \and [\domainstate[i],g_i = \semmakereg{\domainname} (\domainstate[i-1],\reg{r_i}) \wedge \registermap[i] = \registermap[i-1][\reg{r_i} \mapsto g_i]  ]_{i=1\dots m}  \and [\domainstate[m+i+1] = \semset{\ty{ty_i}}{\domainname} (\domainstate[m+i], \registermap[m](\reg{p_i}),v_i)]_{i=0\dots n}
}
{\domainstate[0], \module\vdash (s,\stackk)
  \xrightarrow{\instr} ((\lodinfunction',\lllabel{init},\lllabel{init},0,\registermap[m],\emptyset)),s':\stackk
} 
    }
}

\newcommand{\retvoidrule}{\ensuremath{
    \transrule[Return Void]{
  s = (\lodinfunction,\prev,\cur,\pc,\registermap,\frees),\\
  \frees = \{f_1,f_2,\dots, f_n\} \\
  \syntaxeq{\instr}{\llretvoid} \\
  \stackk = s':\stackk'\\
  \lodinfunction =
    (\funcname{N},\registers,\parameters,\blocklabels,\blocksanon,\blockmap,\ty{ret})
}
{\instr = \blockmap(\cur)[\pc] \and [ \domainstate[i] = \semfree{\domainname}(\domainstate[i-1],f_i)]_{i=1\dots n}
}
{\domainstate[0], \module\vdash (s,\stackk)
  \xrightarrow{\instr} (s',\stackk'),\domainstate[n]}
    }
}

\newcommand{\retvalrule}{\ensuremath{
    \transrule[Return Value]{
  s = (\lodinfunction,\prev,\cur,\pc,\registermap,\frees),\\
  \frees = \{f_1,f_2,\dots, f_n\} \\
  \syntaxeq{\instr}{\llret[ty]{\%val}} \\
  \stackk = s':\stackk' \\
  \lodinfunction =
    (\funcname{N},\registers,\parameters,\blocklabels,\blocksanon,\blockmap,\ty{ty})\\
  s' = (\lodinfunction',\prev',\cur',\pc',\registermap',\frees') \\
  \lodinfunction' = (\funcname{func}',\registers',\parameters',\blocklabels',\blocksanon',\blockmap',\ty{ret}'),\\
  \syntaxeq{\instr[c]}{\reg{\%res}\,\defeq\, \op{call}\, \ty{ty}\, \reg{@N}\, (\ty{ty1}\, \reg{\%\hat p1} \dots \ty{tyn}\, \reg{\%\hat pn})} \\
  \blockmap'(\pc'-1) = \instr[c] \\
  \regvar[v] = \registermap'(\reg{\%res}) 
}
{\instr = \blockmap(\cur)[\pc] \and [ \domainstate[i] = \semfree{\domainname}(\domainstate[i-1],f_i)]_{i=1\dots n}
}
{\domainstate[0], \module\vdash (s,\stackk)
  \xrightarrow{\instr} (s',\stackk'),\semset{\ty{ty}}{\domainname}(\domainstate[n],\regvar[v],\semevalreg[\ty{ty}]{\domainname}(\domainstate[0],\registermap(\reg{\%val})))}
    }
}

\newcommand{\phirule}{\ensuremath{
    \transrule[Phi]{
  s = (\lodinfunction,\prev,\cur,\pc,\registermap,\frees),\\
  s' = (\lodinfunction,\prev,\cur,\pc+1,\registermap,\frees),\\
  \lodinfunction = (\funcname{func},\registers,\parameters,\blocklabels,\blocksanon,\blockmap,\ty{ret})\\
  \syntaxeq{\instr}{\reg{\%res} \defeq \llphiname[ty] \llphiopt{inp1}{lab1} \dots \llphiopt{inpn}{labn}}\\
  \exists i, \lllabel{labi} = \prev\\
  \regvar[{inp}] = \registermap(\reg{\%inpi})\\
  \regvar[{res}] = \registermap(\reg{\%res})
}
{\instr = \blockmap(\cur)[\pc] \and \blockmap(\cur)[\pc+1]\in\phiinst{\registers}{\blocklabels} \and\domainstate[0],\module\vdash (s',\stackk) \xrightarrow{\instr[1]} (s'',\stackk),\domainstate[1]}
{ \domainstate[0], \module\vdash (s,\stackk)
  \xrightarrow{\instr} (s'',\stackk),\semset{\ty{ty}}{\domainname} (\domainstate[1],\regvar[{res}],\semevalreg{\domainname} (\domainstate[0],\regvar[{inp}]))}

    }
}

\newcommand{\phiruletwo}{\ensuremath{
    \transrule[Phi2]{
  s = (\lodinfunction,\prev,\cur,\pc,\registermap,\frees),\\
  s' = (\lodinfunction,\prev,\cur,\pc+1,\registermap,\frees),\\
  \lodinfunction = (\funcname{func},\registers,\parameters,\blocklabels,\blocksanon,\blockmap,\ty{ret})\\
  \syntaxeq{\instr}{\reg{\%res} \defeq \llphiname[ty] \llphiopt{inp1}{lab1} \dots \llphiopt{inpn}{labn}}\\
  \exists i, \lllabel{labi} = \prev\\
  \regvar[{inp}] = \registermap(\reg{\%inpi})\\
  \regvar[{res}] = \registermap(\reg{\%res})
}
{\instr = \blockmap(\cur)[\pc] \and \blockmap(\cur)[\pc+1]\notin\phiinst{\registers}{\blocklabels}}
{ \domainstate[0], \module\vdash (s,\stackk)
  \xrightarrow{\instr} (s',\stackk),\semset{\ty{ty}}{\domainname} (\domainstate[0],\regvar[{res}],\semevalreg{\domainname} (\domainstate[0],\regvar[{inp}]))}
  }
}

\newcommand{\nondetrule}{\ensuremath{
    \transrule[NonDet]{s = 
    (\lodinfunction,\prev,\cur,\pc,\registermap,\frees)\\ \lodinfunction = (\funcname{N},\registers,\parameters,\blocklabels,\blocksanon,\blockmap,\ty{ret}), \\\
     \syntaxeq{\instr}{\llnondet{\%res}{ty}} \\
    \regvar[res] = \registermap(\reg{\%res})
    }{\instr = \blockmap(\cur)[\pc] \and
      \semnondet{\ty{ty}}{\domainname}(\domainstate)=V,\domainstate' \and v\in V }{\domainstate, \module\vdash (s,\stackk)
  \xrightarrow{\instr}
  ((\lodinfunction,\prev,\cur,\pc+1,\registermap,\frees),\stackk),\semset{\ty{ty}}{\domainname} (\domainstate',\regvar[{res}],v)}
}
}

\newcommand{\loadrule}{\ensuremath{
    \transrule[Load]{
      s = 
    (\lodinfunction,\prev,\cur,\pc,\registermap,\frees)\\
    \lodinfunction =
    (\funcname{N},\registers,\parameters,\blocklabels,\blocksanon,\blockmap,\ty{ret}),
    \\
    \syntaxeq{\instr}{\llload[ty]{\%res}{\%add}}\\ 
    \regvar[{addr}] = \registermap(\reg{\%addr}) \\
    \regvar[{res}] = \registermap(\reg{\%res})
  }
  {\instr = \blockmap(\cur)[\pc] \and
      \val\in \semload{\ty{ty}}{\domainname}(\domainstate,\semevalreg[\ptr{ty}]{\domainname}(\domainstate,\regvar[{addr}]))
  }
    {\domainstate, \module\vdash (s,\stackk)
  \xrightarrow{\instr}
  ((\lodinfunction,\prev,\cur,\pc+1,\registermap,\frees),\stackk),\semset{\ty{ty}}{\domainname}
  (\domainstate,\regvar[{res}],\val)}
}}

\newcommand{\storerule}{\ensuremath{
        \transrule[Store]{
  s = 
    (\lodinfunction,\prev,\cur,\pc,\registermap,\frees)\\
    \lodinfunction =
    (\funcname{N},\registers,\parameters,\blocklabels,\blocksanon,\blockmap,\ty{ret}),
    \\
    \syntaxeq{\instr}{\llstore[ty]{\%res}{\%val}{\%addr}}\\
    \regvar[val] = \registermap(\reg{\%val})\\
    \regvar[{addr}] = \registermap(\reg{\%addr})
  }
  {\instr = \blockmap(\cur)[\pc] \and
    \semstore{\ty{ty}}{\domainname}(\domainstate,\semevalreg[\ty{ty}]{\domainname}(\domainstate,\regvar[{val}]),\semevalreg[\ptr{ty}]{\domainname} (\domainstate,\regvar[{addr}])) = \domainstate'
    }
    {\domainstate, \module\vdash (s,\stackk)
  \xrightarrow{\instr}
  ((\lodinfunction,\prev,\cur,\pc+1,\registermap,\frees),\stackk),\domainstate'}
}}

\newcommand{\geprule}{\ensuremath{
    \transrule[GEP]{
  s = (\lodinfunction,\prev,\cur,\pc,\registermap,\frees),\\
  s' = (\lodinfunction,\prev,\cur,\pc+1,\registermap,\frees),\\
  \lodinfunction = (\funcname{func},\registers,\parameters,\blocklabels,\blocksanon,\blockmap,\ty{ret})\\
  \syntaxeq{\instr}{\reg{\%res} \defeq \llgepname{ptr} \llgepind{ty1}{ind1}\dots \llgepind{tyn}{indn}}\\
  \regvar[{ptr}] = \registermap(\reg{\%ptr})\\
  \regvar[{res}] = \registermap(\reg{\%res}) \\
  k = T_{\llnum{ind2},\dots,\llnum{indn}}(\ty{ty})+\llnum{ind1}\cdot\bytesize(\ty{ty})
}
{\instr = \blockmap(\cur)[\pc]  \and\domainstate[0],\module\vdash (s',\stackk) \xrightarrow{\instr[1]} (s'',\stackk),\domainstate[1]}
{ \domainstate[0], \module\vdash (s,\stackk)
  \xrightarrow{\instr} (s'',\stackk),\semset{\ty{ty}}{\domainname} (\domainstate[1],\regvar[{res}],\semptradd(\semevalreg[\ptr{\ty{ty}}]{\domainname} (\domainstate[0],\regvar[{ptr}])),k)}

    }}
{\bf}{\it}
\newtheorem{remark}{Remark}{\bf}{\it}
\newtheorem{definition}{Definition}{\bf}{\it}
\newtheorem{example}{Example}{\bf}{\it}
{\it}{\it}

\begin{document}
\maketitle
\begin{abstract}
  In this work we present our work in developing a software verification
tool for \llvm-code - \lodin - that incorporates  both explicit-state model checking,
statistical model checking and symbolic state model checking
algorithms.
\end{abstract}
\section{Introduction}
Formal Methods, in particular Model
Checking~\cite{DBLP:books/daglib/0020348}, have for many years promised
to revolutionise the way we assert software correctness. It has gained
a  large following in the hardware design industry, but has yet to
become mainstream in the software development industry - and
this despite software being used in a large array of safety-critical
components in e.g. cars and air planes. Nowadays, any
non-trivial component of any system is controlled by an embedded
microprocessor with a control program making software
quality assurance more important than ever. Many case
studies have shown that formal methods is a valuable tool - even in
industrial contexts - but most successful applications have  been
conducted by academic researchers exploring formal methods
usefulness. 

One of the reasons that formal methods have not penetrated the
software industry is, that formal methods require a translation of
the source code to a formal model (e.g. Petri Nets or  Automata) and
the analysis conducted on these formal models. This is problematic as
it requires industry engineers to invest quite some effort into
understanding the formal modelling language and its associated
tool. The diagnostic output for formal tools are also hard to
understand without being an expert in formal methods. As a result,
industry quality assurance relies on extensive testing - which will
have to be done even after applying formal methods - and code reviews. Another
complicating factor in applying the above mentioned workflow is, that
sometimes the engineers do not know the source code intimately - parts
of it might have been auto-generated and some of it might be legacy
code. Attempting to translate code one has not developed
to a formal model is very difficult and error-prone.

In summary, the learning curve of formal methods is 
steep thus industry engineers rely on other methods, and translating
code to formal models is very hard and close to impossible. Formal tools are
needed that understand the source code that industry already uses to
easen the usage of formal tools in industry.

Academics have developed  tools accepting pure code as
inputs~\citep{DBLP:journals/sttt/BeyerHJM07,DBLP:conf/cav/BallR01,DBLP:conf/tacas/FalkeMS13,DBLP:conf/cav/Godefroid97}. A major breakthrough was achieved by tools such as
\textsc{Blast}~\citep{DBLP:journals/sttt/BeyerHJM07} and
\textsc{SLAM}~\citep{DBLP:conf/cav/BallR01} based around a
Counter-Example-Guided-Abstraction-Refinement (CEGAR)~\citep{DBLP:journals/jacm/ClarkeGJLV03},
where a program text is explored symbolic  based on a predicate
abstraction of the program. The predicates are continuously refined to make
the abstraction as detailed as needed. Another approach, pioneered by
the tool \cbmc~\citep{DBLP:conf/tacas/KroeningT14}, is bounded model
checking~\citep{DBLP:journals/ac/BiereCCSZ03}. Here the program
transition system is unrolled a number of times ( in practice by
unrolling loops and inlining function call),
and encoded into a constraint system. During encoding the
assertions can be added that has to be true along any
execution (e.g. that a divisor is never zero). If the resulting constraint
system has a solution where an assertion is true, then the system is
not safe. CEGAR and Bounded Model Checking are incomplete, but are
nevertheless both very successful in locating errors. 

Nowadays the more successful software verification tools are
\cbmc~\citep{DBLP:conf/tacas/KroeningT14} (bounded model checker) and
\cpachecker~\cite{DBLP:conf/cav/BeyerK11} (CEGAR-based tool - and
direct successor of \textsc{Blast}). The tools are among
the  dominating tools in Software Verification competitions\footnote{\url{https://sv-comp.sosy-lab.org}}.

\cbmc\ and \cpachecker\ are both
tied to one source language thus major parts of the tools have to be
implemented for each language they want to  support. A better idea may
be to base the analyses on an intermediate format that can capture the
semantics  of many high level languages. One such intermediate
format is \llvm~\citep{DBLP:conf/cgo/LattnerA04} which at least \ref{toollast} tools
are using:
\begin{enumerate}
\item  \llbmc~\citep{DBLP:conf/tacas/FalkeMS13} follows in the footsteps of \cbmc\ and performs
  bounded model checking on \llvm,  
\item \seahorn~\citep{DBLP:conf/cav/GurfinkelKKN15} has the objective
  of making verification platform for \llvm\ code,  it seems to employ
  mostly CEGAR-based approaches,
\item \klee~\cite{DBLP:conf/osdi/CadarDE08} is a symbolic execution engine performing a
  s symbolic exploration of the state space, in order to find good
  test cases for testing,   and
\item \divine~\cite{BBK+17} is an explicit-state model checker for \llvm\ code\label{toollast}. 
\end{enumerate}

Although previously mentioned tools have paved the way for formal
methods entering industry, they are not without flaws. A lot of them
primarily focus on single-threaded programs which is a problem, because 
industry moves to multi core-architecture and verification
thus needs to take interleaving into account. This interleaving is the
cause of the state space space explosion problem - a problem that the
symbolic representation of \llbmc, \cbmc\ and \cpachecker\ cannot
avoid. Although there has been some work in adapting at least \cbmc\
to concurrent code, it is still an open problem how to verify concurrent
programs efficiently. 

In this paper we present the tool \lodin\, a fairly new tool~\citep{DBLP:conf/fm/LegayNPT18}
offering a range of verification techniques for \llvm. For concurrent programs it
implements explicit-state reachability. Realising an
exhaustive state space search will not scale for large programs, it
also implements under-approximate state space searches through
simulation. For single-threaded programs \lodin\ implements
symbolic exploration akin to \cbmc\ and \llbmc. In this way, \lodin\
distinguishes itself from existing tools by implementing several
techniques into a joint framework.

\lodin\ achieves its ability to
implement different techniques through its flexible
architecture. Another  feature of \lodin\ that sets it apart from 
other formal tools is its extensibility through platform plugins: the core of
\lodin\ implements only the bare minimum semantics of \llvm\ and has
no knowledge of the runtime environment of the program. In real-life
programs, the executing program may call into the runtime
environment which \lodin\ must know about in order to provide correct
verification results. The platform plugins serves as a way to
provide these implementations.


\section{LLVM}
Although the focus of this paper is not to describe the \llvm~\citep{DBLP:conf/cgo/LattnerA04} language
itself, we spend some time on presenting a simplified version of
the \llvm\ instruction set and its semantics. The full \llvm\ language
description is available online~\citep{LLVMSem}. The description we
provide is closely linked to the implementation inside \lodin.  

\subsection{Structure of \llvm programs}
An \llvm\ module consists of functions of which some of them may be
\emph{entry point functions} which are starting points for an \llvm\
process. Functions are  divided into
\emph{Basic Blocks} where a Basic Block is a sequence of instructions
executed in a linear fashion. Basic blocks are named by labels, so that
instructions can direct control to the basic block.
Individual instructions within a basic block can be pure
artihmetic operations, memory allocations, memory accesses, function
calls or instructions that passes control to other basic blocks. Basic
blocks are always terminated by the latter class thus these are called
terminator instructions. Operands to the instructions of an \llvm\
program are kept in so-called registers, and a syntactical requirement
for an \llvm\ is that it must be in \emph{single-static-assignment} \ie\ each
register is only assigned once. 

In \autoref{lst:example} is shown a very short \llvm\ program. The
program consists of a single function $\funcname{main}$ (which is also
the entry point) that
consists of three basic blocks $\lllabel{init,blk}$ and
$\lllabel{succ}$.  The blocks covers lines $4-5$, $7-10$ and $12-13$
respectively. The terminating instruction links \lllabel{init} to
block \lllabel{blk} and links \lllabel{blk} to \lllabel{succ} and
\lllabel{blk}. We refer to \autoref{fig:cfgexample} for a graphical depiction of how the basic blocks are linked together. 

\begin{lstllvm}
\begin{lstlisting}o
; Function Attrs: nounwind uwtable
define void @main() #0 {
init:
	br label %blk				
blk:				
	%x = phi i32 [ %z, %blk ], [ 0, %init ]
	%z = phi i32 [ %x, %blk ], [ 1, %init ]
	%b = icmp eq i32 %x, %z
	br i1 %b, label  %succ, label %blk
succ:
	%y = add i32 0,  1
	ret i32 1
}
\end{lstlisting}
  \caption{An example \llvm\ module with a single entry point \funcname{main}.}\label{lst:example}
\end{lstllvm}

\begin{figure}[t]
  \centering
  \begin{tikzpicture}[auto,
  node distance = 12mm,
  start chain = going below,
  box/.style = {draw,rounded corners,blur shadow,fill=white,
        on chain,align=left}]
 \node[box] (b1)    {\lllabel{init:}\\ \llbr{blk}};      
 \node[box] (b2)    {\lllabel{blk:}\\ \reg{\%x} = \llphiname \llphiopt{z}{blk} \llphioptval{0}{init} \\ \reg{\%z} = \llphiname \llphiopt{x}{branch} \llphioptval{1}{init} \\ \lleq[\tyi{32}]{\%b}{\%x}{\%z} \\ \llbrcond{\%b}{succ}{blk}};      
 \node[box] (b3)    {\lllabel{succ:} \\ \lladd{\%y}{0}{1} \\ \llret{1}};     
 \begin{scope}[rounded corners,-latex]
   \path (b1) edge (b2);
   \path (b2) edge[in=190,out=175,loop]  (b2);
   \path (b2) edge (b3);
 \end{scope}
\end{tikzpicture}

\caption{Control Flow Graph of \autoref{lst:example}.}
\label{fig:cfgexample}
\end{figure}
\paragraph{LLVM Types}
All operations in \llvm\ are typed, either with an arbitrary width bitvector, a compound datatype\footnote{Like C-Style structs} or a memory pointer. The bitvector is denoted \tyi{n} where n is the
width.  For our discussion, we restrict ourselves to bitvectors that are multiple of bytes thus we let \[\inttypes = \set{\tyi{n} \mid \tycolo{n}\in\{8,16,24,32,\dots,\} }\] be the set of
all integer types in \llvm. .If $\ty{ty_1},\dots,\ty{ty_n}$ are \llvm\ types
then $\comptype{\ty{ty_0},\dots  \ty{ty_{n-1}}}$ is a compound type.  We denote by
$\comptypes$ all compound \llvm\ types.  For a type $\comptype{\ty{ty_0},\dots,\ty{ty_{n-1}}}$
and sequence of integers $i_1,\dots,i_k$ we let

\begin{align*}
  \typeind_{i_1,\dots,i_k}(\comptype{\ty{ty_1},\dots,\ty{ty_{n-1}}}) &= \typeind_{i_2,\dots,i_k}(\ty{ty_{i_1}}) \\
  \typeind_\epsilon(\ty{ty}) &= \ty{ty},
\end{align*}
A memory pointer type  to a type $\ty{ty}$ is  denoted \ptr{ty}.  \llvm\ leaves the bithwidth of pointer types unspecified -
for the remainder of this paper we assume it is 64 bit. As is customary in C-style languages, \llvm\ includes the
\ty{void} type used to signify a function does not return a
value.

It will
often be convenient to talk about the byte-size of a type. We therefore define the function
\[
  \bytesize(\ty{ty}) =
  \begin{cases}
    \frac{n}{8} \text{ if } \ty{ty} = \tyi{n} \\ 
    \sum_{i=1}^{n} \bytesize ({\ty{ty_i}}) \text{ if } \ty{ty} = \comptype{\ty{ty_1},\dots, \ty{ty_n}} \\
    8 \text{ if } \ty{ty} = \ptr{\tyi{n}} \\
  \end{cases}
\]

We let $\types$ denote the set of all types in \llvm. 

\paragraph{LLVM instructions} Let $\registers$ be a set of registers, $\blocklabels$ be a finite set
of basic block labels and let $\funcnames$ be a finite set of function
names, then \autoref{tab:instructions} displays the instruction set
used in our discussion of \llvm. In the table $\basicinstructions{\registers} =
\arithinst{\registers}\cup\logicinst{\registers}\cup\meminst{\registers}\cup\cmpinst{\registers}\cup\lodininst{\registers}$
are the basic instructions while $\terminst{\registers}{\blocklabels}$ are
instructions terminating a basic blocks (e.g. jumps). A short
description of the intendend meaning of the instruction classes may be
in order:

\begin{description}
\item[\arithinst{\registers}] Instructions in this class are
  arithmetic instructions that takes two registers (\reg{\%inp1} and \reg{\%inp2}, perform the
  mathematical  operation and store the result in \reg{\%res}. It is worth noting that since \llvm\ has no signed and
unsigned types it instead has signed and unsigned versions of some
instructions. Prime examples of this is the remainder 
(\op{rem}) and the division (\op{div}) instructions. Signed and unsigned versions
are distinguished by the prefixes 's' and 'u'. 
\item[\logicinst{\registers}] This class consists of instructions
  performing bitwise operations. It might be worth mentioning
  the bit shift operations. Shifting to the left, \op{shl},
  is performed by moving the bit pattern towards the most significant bit and pad with zeros. For Shifting to
  the right, \llvm\ has to operations \op{lshr} and \op{ashr}. The
  \op{lshr} is similar to left shifting with the difference that
  the pattern is shifted to the least signifant bit and called a
  logical shift. The \op{ashr} is on the other hand a arithmetic
  right shift, which preserves the sign bit of the pattern.
\item[\meminst{\registers}] This instructions class has instructions
  for allocating memory, loading a value from a memory
  address and a value  at a memory addres. A special instruction
  in this class is the \op{getelementptr} instruction indexing into a compound
  type stored in memory. It can be thought of as the dereferencing
  operator in C.  
\item[\cmpinst{\registers}] This class of instructions are used for
  comparing the values of registers. As an example,
  \llule[i32]{\%res}{\%inp1}{\%inp2} compares if \reg{\%inp1} is less
  than or equal to \reg{\%inp2} while interpreting \reg{\%inp1} and 
  \reg{\%inp2} as unsigned integers.
\item[\terminst{\registers}{\blocklabels}] This class consists of
  instructions terminating a block. A terminating action can either be
  a jump to another block or a return from a function. For jumping
  there are two different version:
  The unconditional version \llbr{block} that jumps to the specified
  block no matter what, and the conditional\\ \llbrcond{\%cond}{ttblock}{ffblock} that jumps to
$\lllabel{ttblock}$ if the pattern in $\reg{\%cond}$  corresponds to
true and to  $\lllabel{ffblock}$ otherwise. There are also two return
instructions:  an instruction ($\llretvoid$) that does not return a value
and one that does ($\llret[ty]{\%res}$).

\item[\callinst{\registers}{\funcnames} ]
 Instruction for calling other functions. 
 The nstruction for calling a function with name $\funcname{func}$
is  $\reg{\%res}\,=\, \op{call}\, \ty{ret}\, \reg{@func}\, (\ty{ty1}\, \reg{\%\hat p1} \dots
\ty{tyn}\, \reg{\%\hat pn}).$ As one would expect, this pass control
to the function \funcname{func}, passes $\reg{\%\hat p1}\dots
\reg{\%\hat p1}$ as parameters and stores the result of the function
call into $\reg{\%res}$.
\item[$\phiinst{\registers}{\blocklabels}$] The instruction class
$\phiinst{\registers}{\blocklabels}$ consists of instructions selecting  a value based on which basic block control flowed from. The
instructions are needed, because \llvm-programs are in
single-static-assignment form. The instructions are only allowed in the start of a
basic block and must be executed simultaneously \ie\ the evaluation of
one phi-instruction cannot affect the result of another in the same block. 
\item[\lodininst{\registers}] This class is a set of ``extension
  instructions'' used by \lodin. Currently it only consists of
  instructions that returns a non-deterministic value. 
\end{description}

\begin{remark}
  All instructions in \autoref{tab:instructions} can take constants as
  parameters in addition to real registers. For ease of exposition  we
  will, however, treat constants as standard registers.  
\end{remark}

\begin{table*}[tb]
  \centering
  \begin{tabular}{l l l}
    \toprule
    \multirow{4}{*}{{\arithinst{\registers}}}& \lladd[ty]{\%res}{\%inp1}{\%inp2} &
                                        \llsub[ty]{\%res}{\%inp1}{\%inp2}
    \\
    & \llmul[ty]{\%res}{\%inp1}{\%inp2}  \\
    & \lludiv[ty]{\%res}{\%inp1}{\%inp2} & \llsdiv[ty]{\%res}{\%inp1}{\%inp2}\\
    & \llurem[ty]{\%res}{\%inp1}{\%inp2} & \llsrem[ty]{\%res}{\%inp1}{\%inp2}\\
                                                                     \midrule
    \multirow{3}{*}{{\logicinst{\registers}}}
                                                                     &\llshl[ty]{\%res}{\%inp1}{\%inp2} &
                                        \lllshr[ty]{\%res}{\%inp1}{\%inp2}\\
    &\llashr[ty]{\%res}{\%inp1}{\%inp2} &
                                         \lland[ty]{\%res}{\%inp1}{\%inp2}\\
    &\llor[ty]{\%res}{\%inp1}{\%inp2} &
                                        \llxor[ty]{\%res}{\%inp1}{\%inp2}\\ \midrule
    \multirow{2}{*}{{\meminst{\registers}}}&\llallocaone[ty]{\%res } &
                                                                       \reg{\%res}
                                                                       \defeq
                                                                       \llgepname{ptr}
                                                                       \llgepind{ty1}{ind1}\dots \llgepind{tyn}{indn} 
    \\
                                           &\llload[ty]{\%res}{\%addr} & \llstore[ty]{\%res}{\%val}{\%addr}
    \\\midrule
    \multirow{6}{*}{{\cmpinst{\registers}}}&\lleq[ty]{\%res}{\%inp1}{\%inp2} & \llne[ty]{\%res}{\%inp1}{\%inp2}\\
    &\lluge[ty]{\%res}{\%inp1}{\%inp2} &
                                        \llugt[ty]{\%res}{\%inp1}{\%inp2} \\
    &\llule[ty]{\%res}{\%inp1}{\%inp2} &
                                        \llult[ty]{\%res}{\%inp1}{\%inp2} \\
    &\llsge[ty]{\%res}{\%inp1}{\%inp2} &
                                        \llsgt[ty]{\%res}{\%inp1}{\%inp2} \\
    &\llsle[ty]{\%res}{\%inp1}{\%inp2} &
                                        \llslt[ty]{\%res}{\%inp1}{\%inp2} \\\midrule
    \multirow{2}{*}{{\terminst{\registers}{\blocklabels}}}
                                                                     &
                                                                       \llretvoid
                                                                                                         &
                                                                                                           \llret[ty]{\%res}
    \\
                                                                     & \llbr{block} & \llbrcond{\%cond}{ttblock}{ffblock}\\
    \midrule 
    {\phiinst{\registers}{\blocklabels}} & \reg{\%res} \defeq \llphiname[ty] \llphiopt{inp1}{lab1} \dots \llphiopt{inpn}{labn} \\
    \midrule 
    \callinst{\registers}{\funcnames} & \reg{\%res}\,\defeq\, \op{call}\, \ty{ret}\, \funcname{func}\, (\ty{ty1}\,
    \reg{\%\hat p1} \dots \ty{tyn}\, \reg{\%\hat pn})
    {{\lodininst{\registers}}} & \llnondet{\%res}{ty}\\
    
    \bottomrule
                                         
  \end{tabular}
  \caption{Basic instructions over a set of registers~\registers~and
    basic block names $\blocklabels$, where $\reg{\%cond},\reg{\%res},\reg{\%inp1},\dots,\reg{\%inpn},\in\registers$,  $\lllabel{block},\lllabel{ttblock},\lllabel{ffblock},\lllabel{lab1},\dots,\lllabel{labn}\in \blocklabels$, $\reg{@func} \in \funcnames$ and for all $i$, $\llnum{indi}\in\mathbb{Z}$.} 
  \label{tab:instructions}
\end{table*}

\paragraph{Formal Definitions of LLVM Modules}
In the introduction to this section, we mentioned that LLVM programs
consists of functions (of which some may be program entry points) and
functions consists of  basic blocks. We are now turning towards giving
propert formal definitions of these concepts. 
\begin{lstllvm}
\begin{lstlisting}
define dso_local i32 @main() {
init:  
  %1 = call i32 (...) @__VERIFIER_nondet_int()
  %2 = icmp ne i32 %1, 0
  br i1 %2, label branch, label end
branch:
%4 = add nsw i32 %1, 1
  br label end
end:
  %.0 = phi i32 [ %4, branch ], [ %1, init ]
  ret i32 %.0
}
\end{lstlisting}
  \caption{Example program for using \llphiname}\label{prgm:phiname}
\end{lstllvm}
\begin{definition}[Basic Block]
  Let $\blocklabels$ be a set of labels, $\funcnames$ be a set
  of functions names  and \registers\ be a set of
  registers, then a basic block, $\block$,  is a finite  sequence
  $\inst{0}\inst{1}\dots\inst{n}$ of instruction where
  \begin{itemize}
  \item for all $i < n$, $\inst{i} \in\basicinstructions{\registers}\cup\callinst{\registers}{\funcnames}\cup\phiinst{\registers}{\blocklabels}$,
  \item $\inst{n}\in\terminst{\registers}{\blocklabels}$ and
  \item if $\inst{i}\in\phiinst{\registers}{\blocklabels}$ then $\forall j<i$, $\inst{j}\in \phiinst{\registers}{\blocklabels}$.
  \end{itemize}

  We denote the set of all possible basic blocks over $\blocklabels$,
  $\registers$ and $\funcnames$ by $\blocks{\registers}{\blocklabels}{\funcnames}$
\end{definition}

\noindent As a convention, if $\block = \inst{0}\inst{1}\dots\inst{n}$ is a
basic block then we write $|\block| = n$ for its length and we let
$\block[i] = \inst{i}$. 

\begin{definition}[Function]
  A function $\lodinfunction$ with $n$ paramters over the function names $\funcnames$  is a tuple
  $(\funcname{N},\registers,\parameters,\blocklabels,\blocksanon,\blockmap,\ty{ret})$
  where
  \begin{itemize}
  \item $\funcname{N}\in\funcnames$ is the functions name,
  \item $\registers$ is a set of registers,
  \item $\parameters = \reg{p_1},\dots,\reg{p_n}$  where for all $i$,
    $\reg{p_i}\in\registers$ , is a sequence of registers used as parameters, 
  \item $\blocklabels$ is a finite set of labels with the requirement
    that $\lllabel{init}\in\blocklabels$ ,
  \item $\blocksanon\subseteq \blocks{\registers}{\blocklabels}{\funcnames}$ is a
    finite set of blocks,
  \item $\blockmap : \blocklabels \rightarrow \blocksanon$ assigns
    each block label a basic block and
  \item $\ty{ret} \in \types$ is the return type of the function. 
  \end{itemize}
\end{definition}

\begin{definition}[Program Entry Point]
  A program entry point is a function
  $(\funcname{N},\registers,\emptyset,\blocklabels,\blocksanon,\blockmap,\ty{void})$.
  \label{def:entrypoint}
\end{definition}

\begin{definition}[Module]
  An \llvm\ module \module\ is a tuple $(\functions,\entrypoints)$ where
  \begin{itemize}
  \item $\functions =  \{\lodinfunction[1],\dots\lodinfunction[n]\}$ is a collection of functions where $\forall i, \lodinfunction[i] = 
  (\funcname{N_i},\registers[i],\parameters[i],\blocklabels[i],\blocksanon[i],\blockmap[i],\ty{ret_i})$, and for all $k\neq j$, $\registers[k]\cap\registers[j] = \emptyset$ and  
\item  $\entrypoints = k_1,\dots,k_m$ is a list of indices defining the entry
  functions  i.e. $\forall 1\leq i\leq m , \lodinfunction[i]$ is an entry
  point function.
  \end{itemize}
\end{definition}

For module $\module = (\functions,\entrypoints)$ we abuse notation slightly and allows
writing $\lodinfunction\in \module$ whenever $\lodinfunction\in \functions$.

\paragraph{Well-typedness}
For each register in $\reg{\%r} \in \registers$ we assign a type from $\ty{t}\in\types$ and
write $\reg{\%r} \typej \ty{t}$ to denote that \reg{\%r} has type
\ty{t}. If a list of registers $\reg{\%1},\dots,\reg{\%n}$ has the
same type $\ty{ty}$, we write $\reg{\%1},\dots,\reg{\%n} \typej
\ty{ty}$. Generalising this notation to an instruction \instr, we
write $\instr \typej \ty{ty}$ to denote $\instr$ is well-typed with
type $\ty{ty}$. \autoref{fig:typerules} shows the type rules of \llvm\  
instructions. 
\begin{figure*}[tb]
  \centering
  \begin{mathpar}
    \typerule{Binary}{\ty{ty}\in\inttypes \and\reg{\%res},\reg{\%inp1},\reg{\%inp2} \typej
      \ty{ty}\and }{(\llbin{\%res}{inst}{ty}{\%inp1}{\%inp2}) \typej\ty{ty}} \and
    \typerule{Compare}{\reg{\%res} \typej \tyi{8} \and
      \reg{\%inp1},\reg{\%inp2} \typej
      \ty{ty} \and \ty{ty}\in\inttypes}{(\llcmp[cc]{ty}{\%res}{\%inp1}{\%inp2}) \typej  \tyi{8}} \and
    \typerule{Alloca}{\reg{\%res} \typej
      \ptr{ty}}{\llallocaone[ty]{\%res}\typej \ptr{ty}} \and
    \typerule{Load}{\reg{\%addr} \typej \ptr{ty} \and \reg{\%res}
      \typej \ty{ty}}{\llload[ty]{\%res}{\%addr} \typej \ty{ty}} \and
    \typerule{Store}{\reg{\%val}\typej \ty{ty} \and \reg{\%addr}
      \typej \ptr{ty} }{\llstore[ty]{\%res}{\%val}{\%addr} \typej
      \ty{void}}
    \and
    \typerule{Phi}{\reg{\%res},\reg{\%lab1},\dots,\reg{\%regn} \typej \ty{ty}}{\reg{\%res} \defeq \llphiname[ty]
      \llphiopt{inp1}{lab1} \dots \llphiopt{inpn}{labn} \typej
      \ty{ty}} \and 
    \typerule{Ret1}{}{\llretvoid} \and
    \typerule{Ret2}{\reg{\%res} \typej \ty{ty}}{\llret[ty]{\%res} \and 
      \typej \ty{ty}} \and
    \typerule{Branch1}{}{\llbr{block} \typej \ty{void}} \and
    \typerule{Branch2}{\reg{\%cond} \typej \tyi{8}}{\llbrcond{\%cond}{ttblock}{ffblock} \typej \ty{void}}
    \and
    \typerule{NonDet}{\reg{\%res} \typej \ty{ty}}{\llnondet{\%res}{ty}
      \typej \ty{ty}}
    \and
    \typerule{Call}{\reg{\%res} \typej \ty{ret} \and
    \big[
    \reg{\% pi},\reg{\%\hat pi} \typej \ty{tyi}
    \big]_{i=1\dots n}}{\reg{\%res}\,\defeq\, \op{call}\, \ty{ret}\,
    \reg{@func}\, (\ty{ty1}\, \reg{\%\hat p1} \dots \ty{tyn}\,
    \reg{\%\hat pn})} \and
  \typerule{GEP}{\reg{\%res} \typej
    \ptr{res} \and\ty{res} = \typeind_{\llnum{ind_2}\dots\llnum{ind_n}}(\ty{ty})}{\reg{\%res}
    \defeq  \llgepname{ptr} \llgepind{ty1}{ind1}\dots
    \llgepind{tyn}{indn} \typej \ptr{res}}
    \end{mathpar}
    \caption{Type rules for \llvm\ for which we have
      $(\llbin{\%res}{inst}{ty}{\%inp1}{\%inp2}\in\arithinst{\registers}\cup\logicinst{\registers}$
    and $(\llcmp[cc]{ty}{\%res}{\%inp1}{\%inp2}) \in \cmpinst{\registers}$}\label{fig:typerules}
\end{figure*}
 For a function $\lodinfunction = (\funcname{N},\registers,\parameters,\blocklabels,\blocksanon,\blockmap,\ty{retty})$ we write $\returnInsts(\lodinfunction)$ to get
all return instructions within that functions basic blocks. Given this
we say that $\lodinfunction$ is well-typed $(\lodinfunction \typej \ty{retty})$ if
for all $\instr\in\returnInsts(\lodinfunction)$, $\instr\typej \ty{retty}$
and all other instructions are well-typed. 

\paragraph{Modelling External Dependencies}
A common problem in software verification is that the system we want
to verify depends on external library functions (\eg \libname{libc}), or functions interacting
directly with the operating system (\eg \libname{pthread}). In
principle we could extend the \llvm\ language with implementations for
all these external function calls but it would unnecessarily inflate
the semantics, and the semantics would have to be redefined  for each
external library and operating system.

\lodin\ combats this problem in two ways:
\begin{inparaenum}
\item \lodin\ extends the \llvm\ language with the \llnondet{\%1}{ty}\
  instruction that returns non-deterministic values, allowing
  a programmer to replace external function calls with \llnondet{\%1}{ty} and
  thereby explore all possible results of external function calls, and
\item \lodin\ allows programmers to extend the \lodin\ interpreter through platform
  plugins that provide implementations  of external functions.  Calls
  to external function calls are syntactically indistinguishable from function defined in the \llvm\
  module itself.    
\end{inparaenum}


\subsection{Contextual Interface}
\label{sec:semantics}
\lodin\ has been developed with reusability in mind allowing to use core components for both explicit state analysis and
symbolic state analysis. The semantics we present in the following
reflect this reusability by defining the core semantics in terms of a \emph{context}. The context is responsible for
representing the register values, how memory is represented and for
implementing  operations on registers. The core semantics ``just''
translate the \llvm\ instruction set to operations on context states
and keeps track of the control flow. In
some sense one could consider the context being a ``virtual machine''.

A context provides the \llvm\
program with an infinite set of  register variables which the
context maps to actual values. The intention is that a \llvm\ program
maps \llvm\ registers to context register variables  i.e. uses a
redirection table to obtain the values of the \llvm\ registers. This
does end up complicating the semantics slightly, but allows calling
a function twice in the \llvm\ program \ie\ enables recursion.

\begin{definition}[Context]
  A context is a tuple $\domainname =
  (\domainstates,\initdomainstate,\domainfortypefname{\domainname},\regvars,\semfalse{\domainname})$ where
  \begin{itemize}
  \item $\domainstates$ is a set of configuration states for the
    context,
  \item $\initdomainstate[\domainname]\in\domainstates$ is the initial context state,
  \item $\domainfortypefname{\domainname}$ assigns to each $\ty{ty}\in\types$ a range
    of values that type can attain values within,
  \item $\regvars$ is an infinite set of register variables,
  \item $\semfalse{\domainname}\in\domainfortype{\tyi{8}}$ is a representation for ``false'' .
  \end{itemize}
\end{definition}

A collection of operations are needed for a \llvm\ program to  manipulate the states of a context. Most
of these operations are just semantical functions for \llvm\
instructions (see Table~\ref{tab:seminstr}). Instead of writing $\circ(S,t_1,t_2) = R$ when applying an
operator, we use an infix notation $\apply{t_1 \circ t_2}{S} =
R$. Besides the instructions in \autoref{tab:seminstr} we need
instructions for creating new register variables (\semmakereg), evaluate the value of a register variable
(\semevalreg[\ty{ty}]{\domainname}), loading (\semload{\ty{ty}}{\domainname})
and storing (\semstore{\ty{ty}}{\domainname}) values from/to memory, allocating memory
(\semalloc{\ty{ty}}{\domainname}) and free'ing memory (\semfree). We discuss them briefly in the following from a usage-perspetice:
\begin{table*}[tb]
  \centering
  \begin{tabular}{l l l l}
    &  Instruction & Operator  &  Signature \\
    Addition & \op{add} &\semadd{\ty{ty}}{\domainname} &
                    $\domainstates\times\domainfortype{\ty{ty}}\times\domainfortype{\ty{ty}}\rightarrow
                    2^{\domainfortype{\ty{ty}}}$\\
    Subtraction &  \op{sub} & \semsub{\ty{ty}}{\domainname} &
                             $\domainstates\times\domainfortype{\ty{ty}}\times\domainfortype{\ty{ty}}\rightarrow
                    2^{\domainfortype{\ty{ty}}}$\\
    Multiplication & \op{mul} &\semmul{\ty{ty}}{\domainname} &
                             $\domainstates\times\domainfortype{\ty{ty}}\times\domainfortype{\ty{ty}}\rightarrow
                    2^{\domainfortype{\ty{ty}}}$\\
    Unsigned Division & \op{div} &\semudiv{\ty{ty}}{\domainname} &
                             $\domainstates\times\domainfortype{\ty{ty}}\times\domainfortype{\ty{ty}}\rightarrow
                    2^{\domainfortype{\ty{ty}}}$\\
    Signed Division & \op{sdiv} &\semsdiv{\ty{ty}}{\domainname} &
                                                                  $\domainstates\times\domainfortype{\ty{ty}}\times\domainfortype{\ty{ty}}\rightarrow
                                          2^{\domainfortype{\ty{ty}}}$\\
    Signed Remainder & \op{rem} &\semsrem{\ty{ty}}{\domainname} &
                             $\domainstates\times\domainfortype{\ty{ty}}\times\domainfortype{\ty{ty}}\rightarrow
                                        2^{\domainfortype{\ty{ty}}}$\\
    Unsigned Modulo & \op{srem} &\semurem{\ty{ty}}{\domainname} &
                             $\domainstates\times\domainfortype{\ty{ty}}\times\domainfortype{\ty{ty}}\rightarrow
                    2^{\domainfortype{\ty{ty}}}$\\
    Shift left & \op{shl} &\semshl{\ty{ty}}{\domainname} &
                             $\domainstates\times\domainfortype{\ty{ty}}\times\domainfortype{\ty{ty}}\rightarrow
                    2^{\domainfortype{\ty{ty}}}$\\
    Logical Shift right & \op{lshr} & \semashr{\ty{ty}}{\domainname} &
                             $\domainstates\times\domainfortype{\ty{ty}}\times\domainfortype{\ty{ty}}\rightarrow
                    2^{\domainfortype{\ty{ty}}}$\\
    Arithmetic shift right & \op{ashr} &\semashr{\ty{ty}}{\domainname} &
                                        $\domainstates\times\domainfortype{\ty{ty}}\times\domainfortype{\ty{ty}}\rightarrow
                    2^{\domainfortype{\ty{ty}}}$\\
    Bitwise and & \op{and} &\semand{\ty{ty}}{\domainname} &
                                        $\domainstates\times\domainfortype{\ty{ty}}\times\domainfortype{\ty{ty}}\rightarrow
                    2^{\domainfortype{\ty{ty}}}$\\
    Bitwise or & \op{or} &\semor{\ty{ty}}{\domainname} &
                                        $\domainstates\times\domainfortype{\ty{ty}}\times\domainfortype{\ty{ty}}\rightarrow
                    2^{\domainfortype{\ty{ty}}}$\\
    Bitwise xor & \op{xor} &\semand{\ty{ty}}{\domainname} &
                                        $\domainstates\times\domainfortype{\ty{ty}}\times\domainfortype{\ty{ty}}\rightarrow
                    2^{\domainfortype{\ty{ty}}}$\\
    
    Equality & \op{cmp} \subop{eq} &\semeq{\ty{ty}}{\domainname} &
                                 $\domainstates\times\domainfortype{\ty{ty}}\times\domainfortype{\ty{ty}}\rightarrow
                                 \domainstates\times\domainfortype{\tyi{8}}\times\{\top,\bot\}$ \\
    Non-equality & \op{cmp} \subop{ne} &\semne{\ty{ty}}{\domainname} &
                                 $\domainstates\times\domainfortype{\ty{ty}}\times\domainfortype{\ty{ty}}\rightarrow
                                 \domainstates\times\domainfortype{\tyi{8}}\times\{\top,\bot\}$
    \\
    Signed Greater than & \op{cmp} \subop{sgt} &\semsgt{\ty{ty}}{\domainname} &
                                 $\domainstates\times\domainfortype{\ty{ty}}\times\domainfortype{\ty{ty}}\rightarrow
                                 \domainstates\times\domainfortype{\tyi{8}}\times\{\top,\bot\}$ \\
    Signed Greater than or equal & \op{cmp} \subop{sge} &\semsgt{\ty{ty}}{\domainname} &
                                 $\domainstates\times\domainfortype{\ty{ty}}\times\domainfortype{\ty{ty}}\rightarrow
                                 \domainstates\times\domainfortype{\tyi{8}}\times\{\top,\bot\}$ \\
    Signed Lessr than or equal & \op{cmp} \subop{sle} &\semsle{\ty{ty}}{\domainname} &
                                 $\domainstates\times\domainfortype{\ty{ty}}\times\domainfortype{\ty{ty}}\rightarrow
                                 \domainstates\times\domainfortype{\tyi{8}}\times\{\top,\bot\}$ \\
    Signed Less than & \op{cmp} \subop{slt} &\semslt{\ty{ty}}{\domainname} &
                                 $\domainstates\times\domainfortype{\ty{ty}}\times\domainfortype{\ty{ty}}\rightarrow
                                 \domainstates\times\domainfortype{\tyi{8}}\times\{\top,\bot\}$ \\
    Unsigned Greater than & \op{cmp} \subop{ugt} &\semugt{\ty{ty}}{\domainname} &
                                 $\domainstates\times\domainfortype{\ty{ty}}\times\domainfortype{\ty{ty}}\rightarrow
                                 \domainstates\times\domainfortype{\tyi{8}}\times\{\top,\bot\}$ \\
    Unsigned Greater than or equal & \op{cmp} \subop{uge} &\semugt{\ty{ty}}{\domainname} &
                                 $\domainstates\times\domainfortype{\tyi{8}}\times\domainfortype{\ty{ty}}\rightarrow
                                                                                           \domainstates\times\domainfortype{\ty{ty}}\times\{\top,\bot\}$ \\
    Unsigned Less than or equal & \op{cmp} \subop{ule} &\semule{\ty{ty}}{\domainname} &
                                 $\domainstates\times\domainfortype{\ty{ty}}\times\domainfortype{\ty{ty}}\rightarrow
                                 \domainstates\times\domainfortype{\tyi{8}}\times\{\top,\bot\}$ \\
    Unsigned Less than & \op{cmp} \subop{ult} &\semult{\ty{ty}}{\domainname} &
                                 $\domainstates\times\domainfortype{\ty{ty}}\times\domainfortype{\ty{ty}}\rightarrow
                                 \domainstates\times\domainfortype{\tyi{8}}\times\{\top,\bot\}$
  \end{tabular}
  \caption{Operations for a context $\domainname =
  (\domainstates,\initdomainstate,\domainfortypefname{\domainname},\regvars)$. They
each take as input a context state and operands and returns a new
contet states and a return value. The compare instructions also return
a value in $\{\top,\bot\}$.}\label{tab:seminstr}
\end{table*}

\paragraph{$ \semmakereg{\domainname} : \domainstates[\domainname] \times \registers \rightarrow \domainstates[\domainname] \times \regvars$ }
This function takes a context state $\domainstate[\domainname]$ and a register
$\reg{\%r}$, where $\reg{\%r} \typej \ty{ty}$. It returns a register
variable $\regvar \in \regvars$ that can be used to store values of
$\ty{ty}$ and a new context state
$\domainstate$. Naturally, the context must ensure   
that the register variable $\regvar$ is not already used in
$\domainstate[\domainname]$. 

\paragraph{$\semevalreg{\domainname}  : \domainstates[\domainname] \times \regvars\rightarrow \domainfortype{\ty{ty}}$}
This function takes a context state $\domainstate$ and register variable
$\regvar\in\regvars$, and returns a value in $\domainfortype{\ty{ty}}$. 

\paragraph{$\semset{\ty{ty}}{\domainname} : \domainstates[\domainname] \times \regvars\times \domainfortype{\ty{ty}} \rightarrow \domainstates[\domainname]$}
This function takes a context state $\domainstate$, register variable
$\regvar\in\regvars$ with type $\ty{ty}$ and a value
$v\in\domainfortype{\ty{ty}}$. It returns a new context state
$\domainstate'$ with $\regvar$ bound to the value $v$.

\paragraph{$\semload{\ty{ty}}{\domainname} : \domainstates[\domainname] \times \domainfortype{\ptr{ty}} \rightarrow 2^{\domainfortype{\ty{ty}}}$}
This function takes a context state $\domainstate$ and a memory address in $\domainfortype {\ptr{ty}}$ and returns a subset of $\domainfortype{\ty{ty}}$. 

\paragraph{$\semstore{\ty{ty}}{\domainname} : \domainstates[\domainname] \times \domainfortype{\ty{ty}}\times\domainfortype{\ptr{ty}} \rightarrow \domainstates[\domainname]$}
This function takes a context state $\domainstate$ and  values
$v\in\domainfortype{\ty{ty}}$ and $a\in\domainfortype{\ptr{ty}}$
. It returns a new $\domainstate'$ where the
value the memory address $a$ has been updated to the value $v$.

\paragraph{$\semalloc{\ty{ty}}{\domainname} : \domainstates[\domainname] \rightarrow \domainstates[\domainname] \times \domainfortype{\ptr{ty}}$}
This function takes a context state $\domainstate$  and returns a
tuple $(\domainstate[1]',t_1)$ where $t_1\in\ptr{ty}$ is a newly allocated
memory address with space for a type $\ty{ty}$, and $\domainstate[1]'$ is a
new context state updated with information that $t$ is no longer free
for allocation.

\paragraph{$\semfree{\domainname} : \domainstates[\domainname] \times \bigcup_{\ty{ty}\in\types}\domainfortype{\ptr{\ty{ty}}} \rightarrow \domainstates[\domainname]$}
This function takes a context state $\domainstate$ and a value in $k
\in \bigcup_{i\in\mathbb{B}}\domainfortype{\ptr{\tyi{i}}}$. It returns
a new context state $\domainstate'$ where the memory pointed to by $k$
has been released. 

\paragraph{$\semnondet{\ty{ty}}{\domainname} : \domainstates[\domainname] \rightarrow \domainstates[\domainname] \times 2^{\domainfortype{\ty{ty}}}$}
This function takes a context state $\domainstate$ and returns a
subset of $\domainfortype{\ty{ty}}$ and a new context state. 

\paragraph{$\semptradd{\ptr{ty}}{\domainname} : \domainfortype{\ptr{ty}}\times\mathbb{Z} \rightarrow \domainfortype{\ptr{ty}}$}
This function takes a pointer $p$  and natural number $b$  and returns
a pointer new pointer after adding $b$ bytes to $p$. 

\subsubsection*{Core Semantics}
We are now ready to define the core semantics for a single
\llvm\ process relative to a given context. The state of a
single process (\eg instruction to be executed, what function it is
executing, which block was previously executed, mapping the functions
register to context register variables)  is kept in an
activation record. The activation record also has  a list of memory
addresses, that must be deallocated when control leaves the currently executing
function. If a function calls another function, an activation record
is pushed in front of the current one thus forming a stack of
activation record.

\begin{remark}
  An activation record roughly corresponds to the well-known concept
  of a stackframe. \llvm\ does however not assume the existence of a
  stack and rather in the activation keeps a set of memory addresses
  that must be relased when removing the activation record
  (corresponding to popping the stackframe in stack-based systems). 
\end{remark}

\begin{definition}[Activation Record]
  An activation record, relative to a context
  $(\domainstates,\initdomainstate[\domainname],\domainfortypefname{\domainname},\regvars,\semfalse)$
  is a tuple
  $(\lodinfunction,\prev,\cur,\pc,\registermap,\frees)$
  where
  \begin{itemize}
  \item
    $\lodinfunction =
    (\funcname{N},\registers,\parameters,\blocklabels,\blocksanon,\blockmap,\ty{ret})$
    is the \llvm\ function currently being executed,
  \item $\prev\in\blocklabels$ is the label of the block executed before the current one,
  \item $\cur\in\blocklabels$ is the label of the currently executed basic block,
  \item $\pc \in\mathbb{N}$ is a pointer into the current basic block
    to locate the next instruction to be executed,
  \item $\registermap : \registers \rightarrow \regvars$ maps registers to register variables of the context and
  \item $\frees$ is a set of memory addresses that must be deleted when removing this activation record.
  \end{itemize}
\end{definition}

\begin{remark}
  Intuitively, an activation record is split into two parts:
  \begin{inparaenum}
  \item A \emph{static} part that indicates which instruction to be
    executed, given by $\lodinfunction,\prev,\cur$ and $\pc$, and 
  \item a dynamic part that links the process to the memory model of
    the context, given by $\registermap$ and $\frees$. 
  \end{inparaenum}
\end{remark}

A stack of activation records is a structure $s_1:s_2\dots:s_n$
where each $s_i$ is an activation record. The empty stack is denoted by
$\emptylist$. In the transition rules in
\autoref{fig:memsem}-\ref{fig:transitionrules}, we usually use the notation $s_1:\stackk$
meaning that $s_1$ is the head of the stack and $\stackk$ is the remaining
part of the stack. We also write $\syntaxeq{\instr}{\op{EXPR}}$ to denote that \instr\ is syntactically equivalent to \op{EXPR}. The transition rules  are defined relative to a context state
$\domainstate$ and a module. Given a context state $\domainstate'$ and
module $\module$ the rules define how to execute an instruction
$\instr$ from state  $(s,\stackk)$,
where $s$ is an activation record and $\stackk$ is a stack, to produce the tuple
$((s',\stackk'),\domainstate)$ where $(s',\stackk')$ is a new state and
$\domainstate$ is a new context state. We write this as
\[
  \domainstate,\module \vdash (s,\stackk) \xrightarrow{\instr}
  ((s',\stackk'),\domainstate').
\]

The rules may look intimidating but most of them are fairly
straightforward. As an example let us briefly consider the rule for
binary operators (that are not comparisons) i.e.
\[
  \resizebox{\linewidth}{!}{\binrule}
\]

This rule says, that in order to execute an instruction
$\llbin{\%res}{inst}{ty}{\%inp1}{\%inp2}$ we first figure out which
register variables in $\domainstate$ that contain the values of
$\reg{\%inp1},\reg{\%inp},\reg{\%res}$. This look up is done with calls to
$\registermap$ and results kept in $\regvar[1],\regvar[2],\regvar[res]$. Then we evaluate
the value of $\regvar[1]$ and $\regvar[2]$ in $\domainstate$ via calls to
$\semevalreg{\domainname}$, and the operation corresponding to $\op{inst}$ is
looked up with $\insttoop$ (see \autoref{tab:seminstr} for this mapping)
and applied ( $\apply{\semevalreg{\domainname}(\domainstate,r_1)) \,\insttoop(\op{inst})\,
      \semevalreg(\domainstate,r_2) }{\domainstate}$ )  giving a new
    context state ($\domainstate'$), and the value of the operation ($\val$). $\semset{\ty{ty}}
    (\domainstate',\regvar[res],\val)$ stores this new value in $\regvar[res]$ and
returns the new context state. Finally we update
the program counter $(\pc+1)$.

In the rules special care has to be taken for the
$\llphiname[\ty{ty}]$ instructoins. All of these   must
be evaluated simultaneously. We therefore evaluate the
them in  a big-step fashion where the evaluation of
one instruction also result in evaluating the next instruction (if it is
also a \llphiname[\ty{ty}] instruction). For the \op{getelementptr}
rule, we use the auxillary function
\begin{align*}
T_{i_1,\dots i_n }(\comptype{\ty{ty1},\dots,\ty{tyn}})  &=
  \sum_{k=1}^{i_1-1}\bytesize(\ty{tyk})+T_{i_2,\dots i_n }(\ty{tyi_1}) \\
  T_\epsilon (\ty{ty}) &= 0
\end{align*}
\noindent to calculate the offset needed to access the correct element of the designated type.
\begin{remark}
  If \lodin\ has some functions defined in a platform plugin, the call
  rule in \autoref{fig:transitionrules} is replaced by the
  implementation described in that module instead. Platform functions
  are executed atomically in \lodin.
\end{remark}

\newcommand{\logrulewidth}{.45\textwidth}
\begin{figure}[tbp]
  \centering
  \begin{mathpar}
   \resizebox{.45\textwidth}{!}{\allocrule}
   \and
   \resizebox{.45\textwidth}{!}{\loadrule} \and
   \resizebox{.45\textwidth}{!}{\storerule}  \and
   \resizebox{.45\textwidth}{!}{\geprule}
 \end{mathpar}
  \caption{Transition Rules for memory instructions}
  \label{fig:memsem}
\end{figure}

\begin{figure}[tbp]
  \centering
  \begin{mathpar}
    \resizebox{\logrulewidth}{!}{\brunrule}
    \and
    \resizebox{\logrulewidth}{!}{\brcondtrue} \and
    \resizebox{\logrulewidth}{!}{\brcondfalse} \and
    \resizebox{\logrulewidth}{!}{\retvoidrule}
    \and 
    \resizebox{\logrulewidth}{!}{\retvalrule}
  \end{mathpar}
  \caption{Transition rules for terminator instructions}\label{fig:branchsem}
\end{figure}

\begin{figure}[tbp]
  \centering
  \begin{mathpar}
    \resizebox{\logrulewidth}{!}{
      \phirule
    }
    \and
    \resizebox{\logrulewidth}{!}{
      \phiruletwo
    }
  \end{mathpar}
  \caption{Compare Rules for Phi instructions}\label{fig:phisem}
\end{figure}

\begin{figure}[tbp]
  \centering
  \begin{mathpar}
    \resizebox{\logrulewidth}{!}{\cmpruleT}
  \end{mathpar}
  \caption{Compare Rules for comparison instructions}\label{fig:compsem}
\end{figure}

\renewcommand{\logrulewidth}{.65\linewidth}
\begin{figure*}[tbp]
  \centering
  \begin{mathpar}
    \resizebox{\logrulewidth}{!}{\binrule}
    \and
    \resizebox{\linewidth}{!}{\callrule}
    
    \and
    \resizebox{\logrulewidth}{!}{
      \nondetrule
}
\end{mathpar}
\caption{Miscellaneous rules Rules}\label{fig:transitionrules}
\end{figure*}

\paragraph{Network of Processes}
Let $\module = (\functions,\entrypoints)$ be an \llvm\ module where
$\functions=\{\lodinfunction[1],\dots,\lodinfunction[n]\}$ with $\lodinfunction[i] =
(\funcname{N_i},\registers[i],\parameters[i],\allowbreak\blocklabels[i],\blocksanon[i],\blockmap[i],\ty{ret_i})$
and  $\entrypoints = \{k_1,\dots,k_m\}$ and let $\domainname =
(\domainstates,\initdomainstate,\domainfortypefname{\domainname},\regvars,\semtrue{\domainname},\semfalse{\domainname})$
be a context. We define the transition system $\lts{\domainname}{\module} =
(\netstates,\netinitstate,\systranss{})$ where a state
$\netstate\in\netstates$ is a tuple  $\netstate= (s_1,s_2,\dots,s_m,\domainstate,\module)$ where each $s_i$ is a state of a
process and $\domainstate\in\domainstates$.

A state
$\netstate= (s_1,s_2,\dots,s_i,\allowbreak\dots,s_m,\domainstate,\module)$ may transit
to state $\netstate' = (s_1,s_2,\dots,s_i',\dots,s_n,\allowbreak\domainstate',\module)$ via the
$i^{\text{th}}$ component performing an instruction $\instr$ if $\domainstate,\module \vdash s_i \xrightarrow{\instr} s_i',\domainstate'.$ We write this as $\netstate
\systranss[i]{\instr}\netstate'$.

The initial state $\netinitstate$ is $ ((\actrecord[1],\emptylist),\dots ,
(\actrecord[m],\epsilon),\initdomainstate,\module)$ where $ \actrecord[i] =
(\stub_{\lodinfunction[k_i]},\lllabel{init},\allowbreak\lllabel{init},0,\_,\emptyset)$
and $\stub({\lodinfunction[k_i]})$ is a special stub function shown in \autoref{lst:stub}.

\begin{lstllvm}
\begin{lstlisting}
  define void @stub ()  {
init:
   call void @N ()
   br label %loop
loop:
   br label %loop
   ret void
}
\end{lstlisting}
\caption{Stub function ({$\stub_{\lodinfunction}$} for instantitating an
  entry point $\lodinfunction = (\funcname{N},\registers,\parameters,\blocklabels,\blocksanon,\blockmap,\ty{void})$}\label{lst:stub}
\end{lstllvm}


\section{Representations in \lodin}
In the preceding section we developed the semantics of \llvm\ programs
abstractly \ie we defined an ``interface'' to a context of the
semantics, allowing instantiating different semantics by
modifying the instantiation of this interface. In this section we
develop two instantiations (\domainexplicit and \domainsymbolic) of the interface. The resulting transition semantics for module $\module$,
$\lts{\domainexplicit}{\module}$ (\lts{\domainsymbolic}{\module}), 
we call the explicit (symbolic) semantics. 

\subsection{Explicit Representation}
\label{subsec:expli}
\label{sec:expli}
\paragraph{Bitvectors}
Let $\boolv = \{0,1\}$ then a bitvector of width $n$ is an element
in $\boolv^{n}$. Two special bitvectors are $\zerovec{n} =
(0,0,\dots,0)\in\boolv^{n}$ and $\onevec{n} =
(1,1,\dots,1)\in\boolv^n$.     If $\vec{b} = (b_0,b_2,\dots,b_{n-1})\in \boolv^{n}$ is a
bitvector, then we can access individual bits by indexing into
$\vec{b}$ i.e. $\vec{b}[i] = b_i$. We also allow extracting the sub-vector
$(b_i,\dots,b_j)$ by $\vec{b}[i:j+1]$. If $\vec{b} = 
(b_0,b_2,\dots,b_{n-1})\in \boolv^{n}$, $\vec{c} =
(c_0,\dots,\dots c_{i-1})\in\boolv^{i}$,  $k\in\{0,n-1\}$ and $k+i < n$ then we let
\[
  \vec{b}[k:k+i/\vec{c}] =
  (b_0,b_1,\dots,b_{k-1},c_0,\dots,c_{i-1},b_{k+i},\dots,b_{n-1}). 
\]

Let $\vec{b} = (b_0,b_2,\dots,b_{n-1})\in \boolv^{n}$ be a bitvector,
then we can interpret it as either an unsigned integer or a signed
integer. In the prior case we use the standard binary encoding and
 define $\unsignedBV{\vec{b}} = \sum_{i=0}^n b_i \cdot 2^{i}.$ In the latter case we use 2's-complement encoding and  let
$\signedBV{\vec{b}} = -b_{n-1} 2^{n-1}+\sum_{i=0}^{n-2} b_i 2^{i}.$
To encode a number $n\in\mathbb{N}$  in either binary or 2s-complement
we write $\unencodeBV{n}$ and  $\signencodeBV{n}$ respectively.

The classic bitwise operators, and, or, xor and negation, between
vector  $\vec{b_1},\vec{b_2}\in\boolv^n$  are defined as usual and
denoted  $(\vec{b_1} \bitand \vec{b_2})$, $(\vec{b_1} \bitor
\vec{b_2})$, $(\vec{b_1} \bitxor \vec{b_2})$ and $(\bitneg \vec{b_1})$
respectively. If $\vec{b_1}\in\boolv^n$ is a bitvector,
$d\in\mathbb{N}$ is a number and $d < n$  then we define bit shifting operations as 

\begin{mathpar}
\vec{b_1} \bitlshl  d= \zerovec{n}[d:n/\vec{b_1}[0:n-d]], \and 
\vec{b_1} \bitlshr d = \zerovec{n}[0:d/\vec{b_1}[n-d:n]] \\ \and
\vec{b_1} \bitashr d =
  \begin{cases}
    \zerovec{n}[0:d/\vec{b_1}[n-d:n]] \text{ if  } \vec{b_1}[n-1] = 0 \\
    \onevec{n}[0:d/\vec{b_1}[n-d:n]] \text{ if  } \vec{b_1}[n-1] = 1 \\
  \end{cases}
\end{mathpar}

The $\bitlshl$ ($\bitlshr$) operator is a logic left (right) bitshift
i.e. shift all bits to left (right) and pad with zero. The $\bitashr$
is arithmetic right shift where instead of padding with zero, the bit
vector is padded with the original value of the most significant bit.  

\paragraph{Memory Modelling}
In the explicit semantics we model the memory state of a computer as
a (possibly) infinite length array of memory blocks. Memory blocks are
tagged with their size and the actual content of the block. Formally,
the memory state of program is a function $\memblockmap : \mathbb{N}
\rightarrow
(\mathbb{N}\times(\bigcup_{i\in\mathbb{N}}\boolv^i))\cup\{\memblockunused\})$.
An entry $\memblock(i)$ means that block $i$ of the memory has not
been used. If $\memblock(i) = (k,\vec{b})$ and
$\vec{b}\in\bitvector{k}$ then we say that block $i$ is
\emph{consistent}, has $k$ and $\vec{b}$ is the content of that block. 

\begin{figure}[tb]
  \centering
  \begin{tikzpicture}[remember picture]
   \node (a) {\begin{tabular}{|l|l|}
                
                \multicolumn{2}{c}{$M$} \\
                \hline
                0 & $\bot$ \\ \hline
                1 & $\bot$ \\ \hline
                \multicolumn{2}{c}{\vdots} \\
                \hline
                $\tiny{\mathtt{block}}$ & $(\tiny{\mathtt{size}},$ \tikz[overlay, remember picture] {\node (Inner) {$\cdot$};} $)$ \\
                \hline
                \multicolumn{2}{c}{\vdots} \\
                
              \end{tabular}
            };
            \node[right = of a] (b) {
              \begin{tabular}{|l|l|l|l|l|l|l|l|}
                
                \multicolumn{6}{|c|}{\tiny{\texttt{offset}}} \\
                \hline
                & & &\multicolumn{2}{c}{\dots}  & & & \\
                \hline
                \multicolumn{8}{|c|}{\tiny{\texttt{size}}} \\
              \end{tabular}
            };

            \draw[->] (Inner) -- (b.mid west);
\end{tikzpicture}
\caption{Memory representation in \lodin. Pointers are 64bit integeres split into a 32bit \texttt{base} and a 32bit \texttt{offset}. \lodin\ uses a redirection table ($M$) that store memory blocks, and \texttt{block} indexes into this table, while \texttt{offset} indexes into the memory blocks. The symbol $\bot$ indicates an entry in $M$ is unused.   }
\label{fig:memGraph}
\end{figure}
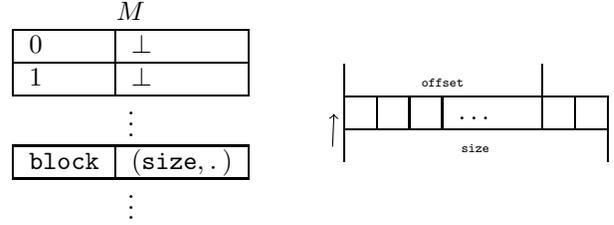
To modify and read from memory, we define the
functions:
\begin{itemize}
\item $\newmem((\memblockmap),i) = (\memblockmap[n\mapsto
  (i,\zerovec{i})]),n$ where
  $n=\mathtt{min}(\{g \mid \memblockmap(g) = \memblockunused\})$,
\item $\free((\memblockmap,\memused),i) = (\memblockmap[i\mapsto
  \memblockunused])$,
\item $\rread(\memblockmap,b,f,len) = \vec{b}[f:f+len]$
  where $\memblockmap(b) = (i,\vec{b})$ and $f+len<i$ and
\item $\wwrite(\memblockmap,b,f,\vec{c},len) =
  (\memblockmap[b\mapsto \vec{b}[f:f+len/\vec{c}],\memused)$
  where $\memblockmap(b) = (i,\vec{b})$ and $f+len<i$ and
\end{itemize}
\begin{figure*}[tbp]
  \centering
  \begin{mathpar}
    \semmakereg{\domainexplicit}((\memblockmap,N,F),\reg{\%r}) = (\memblockmap,N\cup\{i\},F),i \text { where } i = \dmin (\mathbb{N}\setminus N) \and
  \semevalreg{\domainexplicit}((\memblockmap,N,F),i) =
  \begin{cases}
    F(i) & \text{ if } F(i) \in \domainfortype[\domainexplicit]{\ty{ty}}\\
    \mathbf{Error} & \text { Otherwise } \\
  \end{cases}  \and
  \semalloc{\ty{ty}}{\domainexplicit}((\memblockmap,N,F)) = (\memblockmap',N,F),i \text{ if } \ty{ty} = \tyi{n} \wedge  \newmem(\memblockmap,\bytesize(\ty{ty})) = \memblockmap',i   
  \and
  \semfree{\domainexplicit} (((\memblockmap,\memused),N,F),i) =
  \begin{cases}
    (\free((\memblockmap,\memused),k),N,F))& \text{ if }
    \substack{
      i\in\boolv^{64} \\ k = \unsignedBV{i[32:64]}\in \memused \\ \unsignedBV{i[0:32]} = 0
    }\\
    \mathbf{Error} & \text{ otherwise}
  \end{cases}
  \and
  \semload{\ty{ty}}{\domainexplicit}(((\memblockmap,\memused),N,F),i) =
  \begin{cases}
    \{(((\memblockmap,\memused),N,F),\rread((\memblockmap,\memused),k,o,m)) \}& \text { if }  \substack{ k =  \unsignedBV{i[32:64]}\in \memused \\
      \domainfortype[\domainexplicit]{\ty{ty}} = \boolv^{m} \\
      o = \unsignedBV{i[0:32]}*8 \\
      (s,b) = \memblockmap(k) \\ o + m < s } \\
    \{(((\memblockmap,\memused),N,F),g) | g\in\boolv^{m} \} &\text{ if  } \domainfortype[\domainexplicit]{\ty{ty}} = \boolv^{m}\\ 
    \end{cases}
    \and 
    \semstore{\ty{ty}}{\domainexplicit}((((\memblockmap,\memused)),N,F),v,p) =
  \begin{cases}
    ((\wwrite((\memblockmap,\memused),k,o,v,m),N,F)) & \text { if }  \substack{ k =  \unsignedBV{p[32:64]}\in \memused \\
      \domainfortype[\domainexplicit]{\ty{ty}} = \boolv^{m} \\
      o = \unsignedBV{p[0:32]}*8 \\
      (s,b) = \memblockmap(k) \\ o + m < s } \\
      \mathbf{Error} &\text { otherwise}
  \end{cases} \\
\end{mathpar}
 
\caption{Operations for the explicit semantics.}\label{fig:expop1}
\end{figure*}
The initial state of the memory is the function $\memblockmapinit$ where for all $i$, $\memblockmapinit(i) = \memblockunused$.
\par 
Given both a representation of the register values and the memory, we
can now define the explicit context. In the explicit context, we
assign to a type $\ty{in}$ the domain $\boolv^{n}$ and any pointer
type is assigned the domain $\boolv^{64}$. Using a 64-bit bitvector
for representing pointers allows us to use the 32 most significant for indexing into
$\memblockmap$ of the memory and use 32 least significant bits to
index into the actual block.    For a pointer $p\in\boolv^{64}$ we let $\texttt{block}(p) = p[32:64]$ and $\texttt{offset}(p) = p[0:32]$. See \autoref{fig:memGraph} for a graphical depiction of how this work.

\begin{definition}[Explicit Context]
  The explicit context  is the tuple  $\domainexplicit = (\domainstates[\domainexplicit],\initdomainstate[\domainexplicit],\allowbreak\domainfortypefname{\domainexplicit},\mathbb{N},\semfalse{\domainexplicit})$ where
  \begin{itemize}
  \item $\domainstates[\domainexplicit] = \{(\memblockmap,N,F) \mid \memblockmap \text{ is a memory state} \land N\subset \mathbb{N} \land F : \mathbb{N} \rightarrow (\bigcup_{i\in\mathbb{N}} \boolv^i)\cup\{\bot\}\}$
  \item $\initdomainstate[\domainexplicit] = (m_{init}, \emptyset,F)$ where for all $i$, $F(i) = \bot$, 
  \item $\domainfortypefname{\domainexplicit}(\ty{t}) =  \boolv^{8\cdot\bytesize(\ty{t})}$
  \item $\semfalse{\domainexplicit} = \unencodeBV{0}$.
  \end{itemize}
\end{definition}

The operations for modifying the explicit context is provided in 
\autoref{fig:expop1} and \autoref{fig:expop2}. The rules are derived
from the informal description provided at~\citep{LLVMSem}. For the
comparison operators, we give the definition of $\semugt{\ty{ty}}{\domainexplicit}$ and
$\semsgt{\ty{ty}}{\domainexplicit}$ below, and note that the remaining comparison operators are
easily generalised from these. In the rules we let $\semtrue{\domainexplicit}\in\domainfortype[\domainexplicit]{\tyi{8}}$ and require $\semtrue{\domainexplicit} \neq \semfalse{\domainexplicit}$.

\begin{mathpar}
  \semugt{\ty{ty}}{\domainexplicit}(\domainstate,r_1,r_2) =
  \begin{cases}
    (\domainstate,\semtrue{\domainexplicit},\top) & \text{ if } \unsignedBV {r_1} > \unsignedBV
    {r_2},  \\
    (\domainstate,\semtrue{\domainexplicit},\bot)& \text{ otherwise}
  \end{cases}
  \and
  \semsgt{\ty{ty}}{\domainexplicit}(\domainstate,r_1,r_2) =
  \begin{cases}
    (\domainstate,\semtrue{\domainexplicit},\top) & \text{ if } \signedBV {r_1} > \signedBV
    {r_2} \\
    (\domainstate,\semfalse{\domainexplicit},\bot)& \text{ otherwise}
  \end{cases}
\end{mathpar}

\begin{figure*}[tbp]
  \centering
  \begin{mathpar}
    \semadd{\ty{ty}}{\domainexplicit}(S,\vec{b_1},\vec{b_2}) =
  \begin{cases}
    \{(\unencodeBV{b_1+b_2})\} & \text{ if } \substack{ \\ b_i = \unsignedBV{\vec{b_i}}\\b_1+b_2 \leq \unencodeBV{\onevec{m}} } \\
    \{(\unencodeBV{ (b_1+b_2) \% 2^{m}  })  \} &\text{ otherwise }
  \end{cases} \and 
    \semptradd[\domainexplicit](\vec{b_1},k) =\vec{b_2} \text{ where } \mathtt{block}(\vec{b_2}) = \mathtt{block}(\vec{b_1}) \text{ and } \mathtt{offset}(\vec{b_2}) = \mathtt{offset}(\vec{b_1})+k
\and
  \semsub{\ty{ty}}{\domainexplicit}(S,\vec{b_1},\vec{b_2}) =
  \begin{cases}
    \{(\unencodeBV{b_1+b_2})\} & \text{ if } \substack{  b_1 = \unsignedBV{\vec{b_1}}\\
      b_2 = \unsignedBV{\bitneg\vec{b_2}}+1 \\
      b_1+b_2 \leq \unencodeBV{\onevec{m}} } \\
    \{(\unencodeBV{ (b_1+b_2) \% 2^{m}  } )\} &\text{ if } \substack{
      b_1 = \unsignedBV{\vec{b_1} }\\
    b_2 = \unsignedBV{\bitneg\vec{b_2}}+1}
\end{cases}
\and
  \semmul{\ty{ty}}{\domainexplicit}(S,\vec{b_1},\vec{b_2}) =
  \begin{cases}
    \{(\unencodeBV{b_1*b_2})\} & \text{ if } \substack{  b_i = \unsignedBV{\vec{b_i}}\\b_1*b_2 \leq \unencodeBV{\onevec{m}}  \\  b_i = \unsignedBV{\vec{b_i}}} \\
    \{ (\unencodeBV{ (b_1*b_2) \% 2^{m}  })  \} &\text{ otherwise }
  \end{cases}
  \and
  \semudiv{\ty{ty}}{\domainexplicit}(S,\vec{b_1},\vec{b_2}) =
  \begin{cases}
    \{(\unencodeBV{\lfloor b_1 / b_2 \rfloor})\} & \text{ if } \substack{  b_i = \unsignedBV{\vec{b_i}}\\b_2 \neq 0} \\
    \{(\vec{c}) | \vec{c}\in\boolv^{m}   \} &\text{ otherwise }
  \end{cases}
  \and
  \semsdiv{\ty{ty}}{\domainexplicit}(S,\vec{b_1},\vec{b_2}) =
  \begin{cases}
    \{(\signencodeBV{\mathtt{trunc} (b_1 / b_2 )}\} & \text{ if } \substack{  b_i = \signedBV{\vec{b_i}}\\b_2 \neq 0} \\
    \{(\vec{c}) | \vec{c}\in\boolv^{m}   \} &\text{ otherwise}
  \end{cases}
  \and
  \semurem{\ty{ty}}{\domainexplicit}(S,\vec{b_1},\vec{b_2}) =
  \begin{cases}
    \{(\unencodeBV{b1-b2\cdot\lfloor b_1 / b_2 \rfloor})\} & \text{ if } \substack{  b_i = \unsignedBV{\vec{b_i}}\\b_2 \neq 0} \\
    \{(\vec{c}) | \vec{c}\in\boolv^{m}   \} &\text{ otherwise }
  \end{cases}
  \and
  \semsrem{\ty{ty}}{\domainexplicit}(S,\vec{b_1},\vec{b_2}) =
  \begin{cases}
    \{(\signencodeBV{b1- b2\cdot(\mathtt{trunc}( b_1 / b_2))}\} & \text{ if } \substack{  b_i = \signedBV{\vec{b_i}}\\b_2 \neq 0} \\
    \{(\vec{c}) | \vec{c}\in\boolv^{m}   \} &\text{ otherwise}
  \end{cases}
  \and
  \semshl{\ty{ty}}{\domainexplicit}(S,\vec{b_1},\vec{b_2}) =
  \begin{cases}
    \{(\vec{b_1} \bitlshl b_2 )\} & \text{ if } \substack{  b_2 = \unsignedBV{\vec{b_2}} \\ b_2 < m  } \\
    \{(\vec{c} ) | \vec{c}\in\boolv^{m}\} & \text{ otherwise }
  \end{cases}
  \and 
  \semlshr{\ty{ty}}{\domainexplicit}(S,\vec{b_1},\vec{b_2}) =
  \begin{cases}
    \{(\vec{b_1} ) \bitlshr b_2\} & \text{ if } \substack{  b_2 = \unsignedBV{\vec{b_2}} \\ b_2 < m  } \\
    \{(\vec{c}) | \vec{c}\in\boolv^{m}\} & \text{ otherwise }
  \end{cases}
  \and 
  \semashr{\ty{ty}}{\domainexplicit}(S,\vec{b_1},\vec{b_2}) =
  \begin{cases}
    \{(\vec{b_1} \bitashr b_2 )\} & \text{ if } \substack{  b_2 = \unsignedBV{\vec{b_2}} \\ b_2 < m  } \\
    \{(\vec{c}) | \vec{c}\in\boolv^{m}\} & \text{ otherwise }
  \end{cases}
  \and
\semand{\ty{ty}}{\domainexplicit}(S,\vec{b_1},\vec{b_2}) = \{(\vec{b_1} \bitand \vec{b_2} )\}
\and
\semor{\ty{ty}}{\domainexplicit}(S,\vec{b_1},\vec{b_2}) = \{(\vec{b_1} \bitor \vec{b_2} )\}
\and
\semxor{\ty{ty}}{\domainexplicit}(S,\vec{b_1},\vec{b_2}) = \{(\vec{b_1} \bitxor \vec{b_2} )\}
\end{mathpar}
\caption{Operation for the explicit semantics. Throughout these rules
  we assume that $\domainfortype[\explicitdomain] (\ty{ty}) = \boolv^m$, for some $m$. In the rules we use $\mathtt{trunc}$ to denote a rounding operation towards zero.}\label{fig:expop2}
\end{figure*}

\begin{remark}
  Instantiating a model with the explicit context as described so far
  result in a possibly  infinite state space. As a result, an
  exhaustive enumeration of all possible states may not terminate.    
\end{remark}


\subsection{Symbolic Representation}
We have already mentioned that an explicit representation of values in
a program will explode (even without concurrency) in the presence of
non-deterministic values. As an example of this, consider
\autoref{lst:symbneed} which can call the function \funcname{@error}
if and only \reg{\%2} is set to $5$. It is easy for easy for humans
to realise that \funcname{@error} can be called, but a computer with an explicit representation has to enumerate
all $32^2-1$ possible values of \reg{\%2}. 
\begin{lstllvm}
  \begin{lstlisting}
    define void @main() #0 {
init:
  %1 = alloca i32, align 4
  %2 = lodin_nd i32
  store i32 %2, i32* %1, align 4
  %3 = load i32, i32* %1, align 4
  %4 = icmp eq i32 %3, 5
  br i1 %4, label %call, label %done

call:                                      ; preds = %0
  call void (...) @error()
  br label %done
  
  done:                                      ; preds = %5, %0
  ret void
}
\end{lstlisting}
\caption{Example of why Symbolic Representation are necessary}\label{lst:symbneed}
\end{lstllvm}

For combatting this, \lodin\ provides a
symbolic context representation. Instead of representing values
explicitly, the symbolic context gathers all operations performed
during exploration into one large logical formula - known as the path
formula -  that can since be passed to
an SMT-solver. The SMT-solver can  then determine if the formula is
satisfiable and thus if the explored path is feasible.

\subsubsection{Satisfiability Modulo Theories }
An SMT-instance is principally a first order logic formula where some
predicates and functions have special interpretations. These special
interpretations are encapsulated into what is called theories. An
SMT-instance of the theory $\theory$ can be determined to be
satisfiable or not satisfiable by SMT-solver supporting the
$\theory$. We will not invest too much time here in talking about how
SMT-solvers work, but will rather informally discuss the theories we need. 

\paragraph{Theory of Bitvectors}
In the theory of bitvectors, variables are given a  bitvector type
\tyi{n}\footnote{Note we reuse the type name from \llvm}.  The operations that can be performed between bitvectors are
\begin{itemize}
\item the classic bitwise operations, i.e. \bitand, \bitor,
  \bitneg, \bitxor, \bitlshl, \bitlshr\ and \bitashr
\item arithmetic operations (modulo $2^n$),
  i.e. \bitadd, \bitsub, \bitudiv, \bitsdiv, \bitmul, \biturem, \bitsrem\ -
  as in the \llvm\ discussion we need both signed and unsigned
  versions of some operations (indexed by $_u$ and $_s$)
\item comparisons e.g. $=$ and $\leq$,
\item boolean  operations \eg ($\wedge, \vee, \lnot$) 
\item concatenation of bitvectors $\bitconcat$,
\item extraction of subvectors i.e. if $\smtvar$ is a bitvector then $b\smtvar[] [0\dots
  n]$ extract a bitvector with bits $0$ to
  $n-1$. 
\end{itemize}

\begin{remark}
  We reuse the operatorions from our discussin of bitvectors in
  \autoref{subsec:expli}, and require that the SMT-solver implements the
  semantics of the operations as described there. Likewise we write constant bitvectors using the notation from \autoref{subsec:expli}.
\end{remark}

\paragraph{Theory of Arrays}
In this theory an array is a mapping between elements. Elements from
an array can be read using  a \arrselect\ function,  and an
element stored in an array using a \arrstore\ function. We introduce
the array type \arrtype{\tyi{n}}{\tyi{m}} mapping elements from
\tyi{n} to \tyi{m}. If $\smtvar \typej \arrtype{\tyi{n}}{\tyi{m}}$,
$\smtvar[1] \typej \tyi{n}$ and $\smtvar[2] \typej \tyi{m}$ then we
write $\arrstore(\smtvar,\smtvar[1],\smtvar[2])$ to create a new array
that is equal to $\smtvar$ with the only difference that the value of
$\smtvar[1]$ now maps to the value of $\smtvar[2]$.  We also write
$\smtvar[2]  = \arrselect(\smtvar,\smtvar[1])$ to set $\smtvar[2]$
equal to the value kept at position $\smtvar[1]$.

In the following we use $\smtvars[]$ to denote an infinite set of SMT
variables. We also use the restricted sets $\smtvars[\ty{ty}] =
\{\smtvar \in\smtvars \mid \smtvar \typej \ty{ty} \}$. Similarly we
refer by $\smtexprs$ to all SMT expressions over $\smtvars$ and
$\smtexprs[\ty{ty}]$ to all SMT expressions with  type $\ty{ty}$.  

\subsubsection*{The Symbolic Context}
The symbolic context in \lodin\ maps its register variables to SMT
variables and uses a so called path formula to capture all constraints
(assignments and comparisons) encountered during a program
execution. Memory is represented using a SMT array and a SMT variable
points to first place in memory that is free for allocation. 

\newcommand{\used}{\ensuremath{\mathtt{used}}}
\begin{definition}[Symbolic Context]
  The symbolic context for the symbolic semantics is the tuple  $\domainsymbolic = (\domainstates[\domainsymbolic],s_{init},\allowbreak\domainfortypefname{\domainsymbolic},\mathbb{N},\semfalse{\domainsymbolic})$ where
  \begin{itemize}
  \item $\domainstates[\domainsymbolic]$ are tuples
    $(\smtvar[M],\smtvar[f],N,F, \pathform,\used)$ where
    \begin{itemize}
    \item $\smtvar[M] \typej \arrtype{\tyi{64}}{\tyi{8}}$ is an array
      representing the memory state of the program, 
    \item $\smtvar[f] \typej \tyi{64}$ is a pointer into memory 
    \item $N \subseteq \mathbb{N}$ is a set of used register
      variables, 
    \item $ F : \mathbb{N} \rightarrow \smtvars[]\cup\{\bot\}$ 
    \item $\pathform$ is an SMT formula - the path formula - encoding
      the constraints that an explored path has to satisfy, and
    \item $\used\subseteq \smtvars$ is a set of used  SMT variables.
    \end{itemize}
    
  \item $s_{init} = (M,0,F,\semfalse{\domainsymbolic} == \semfalse{\domainsymbolic},\emptyset)$ where for all $n\in\mathbb{N}$ , $F(n) = \bot$
  \item $\domainfortypefname{\domainsymbolic}(\tyi{i}) =
    \smtexprs[\tyi{i}]$, $\domainfortypefname{\domainsymbolic}(\ptr{ty}) = \smtexprs[\tyi{64}]$, and  
    $\domainfortypefname{\domainsymbolic}(\comptype{\ty{ty1},\dots\ty{ty1n}}) = \smtexprs[8\cdot\bytesize(\comptype{\ty{ty1},\dots\ty{ty1n}})]$
  \item $\semfalse{\domainsymbolic} = \zerovec{8}$. 
  \end{itemize}
\end{definition}

The arithmetic instructions (e.g. $\semadd{\ty{ty}}{\domainsymbolic}(\domainstate, \smtvar[1],\smtvar[2]$) that we need to implement for the context is
straightforward to represent. All we need to do is to create an SMT expressions corresponding to the operation,
 Below we give a generalised definition of the rule:

\begin{align*}
  \sim^{\ty{ty}}_{\domainsymbolic} &((\smtvar[M],\smtvar[f],F,
                   \pathform,\used),\smtvar[1],\smtvar[2]) = \smtvar[1]\,\mathtt{SMTOp}\, \smtvar[2]\\
\end{align*}
 For the mapping between $\sim^{\ty{ty}}_{\domainsymbolic}$  and
$\mathtt{SMTOp}$ we 
refer to \autoref{tab:opsmtop}.

The comparison operators are very similar to the binary operator, and
below we provide an example for the
$\semugt{\ty{ty}}{\domainsymbolic}(\domainstate,\smtvar[1],\smtvar[2])$ function
where $\domainstate =  (\smtvar[M],\smtvar[f],F, \pathform,\used)$

\begin{align*}
  \semugt{\ty{ty}}{\domainsymbolic}(\domainstate,\smtvar[1],\smtvar[2])
  = &(\smtvar[M],\smtvar[f],N,F, \pathform\wedge (\smtvar[1] >_u \smtvar[2]), \\
                                                 &\used,\smtvar[1] >_u \smtvar[2],\top)
\end{align*}

For the remainder of the operations we
refer the reader to \autoref{fig:evalsetsymb} and \autoref{fig:loadstoresymb}.
\renewcommand{\arraystretch}{1.5}
\begin{table}[t]
  \centering
  \begin{tabular}{|l|l|}
    \semadd{\ty{ty}}{\domainsymbolic} & \bitadd \\
    \semsub{\ty{ty}}{\domainsymbolic} & \bitsub \\
    \semmul{\ty{ty}}{\domainsymbolic} & \bitmul \\
    \semudiv{\ty{ty}}{\domainsymbolic} & \bitudiv\\
  \end{tabular}\hfill
  \begin{tabular}{|l|l|}
    \semsdiv{\ty{ty}}{\domainsymbolic} & \bitsdiv \\
    \semurem{\ty{ty}}{\domainsymbolic} & \biturem\\
    \semsrem{\ty{ty}}{\domainsymbolic} & \bitsrem\\
    \semshl{\ty{ty}}{\domainsymbolic} & \bitlshl\\
  \end{tabular}\hfill
  \begin{tabular}{|l|l|}
    \semlshr{\ty{ty}}{\domainsymbolic} & \bitlshr\\
    \semashr{\ty{ty}}{\domainsymbolic} & \bitashr\\
    \semand{\ty{ty}}{\domainsymbolic} & \bitand\\
    \semor{\ty{ty}}{\domainsymbolic} & \bitor\\
    \semxor{\ty{ty}}{\domainsymbolic} & \bitxor\\
  \end{tabular}
  \caption{Mapping between semantical operators and SMT operators\label{tab:opsmtop}}
\end{table}

\begin{figure*}
  \centering
  \begin{mathpar}
    \semmakereg{\domainsymbolic}((\smtvar[M],\smtvar[f],N,F,\pathform,\used),\reg{\%r})
    = (\smtvar[M],\smtvar[f],N\cup\{i\},F,\pathform,\used),i \text {
      where } i = \dmin (\mathbb{N}\setminus N) \and

    \semevalreg{\domainsymbolic}((\smtvar[M],\smtvar[f],N,F,\pathform,\used),i)
    =
    \begin{cases}
      F(i) & \text{ if } F(i) \in \domainfortype[\domainsymbolic]{\ty{ty}}\\
      \mathbf{Error} & \text { Otherwise } \\
    \end{cases} \and
    \semset{\ty{ty}}{\domainsymbolic}((\smtvar[M],\smtvar[f],N,F,\pathform,\used),l,\smtvar)
    = \begin{cases}
      (\smtvar[M],\smtvar[f],N,F,\pathform\wedge (F(l) = \smtvar),\used) & \text{ if } l\in N \land \smtvar\in\domainfortypefname{\domainsymbolic}(\ty{ty})\\
      \mathbf{Error} & \text { Otherwise } \\
    \end{cases} \and
    \semalloc{\ty{ty}}{\domainsymbolic}((\smtvar[M],\smtvar[f],N,F,\pathform,\used))
    =
    \begin{cases}
      (\smtvar[M],\smtvar[f'],N,F,\pathform\wedge (\smtvar[f']=\smtvar[f]\bitadd n),\used\setminus\{\smtvar[f]\}),\smtvar[f]& \text{ if } \ty{ty} = \tyi{n}     \\
      (\smtvar[M],\smtvar[f'],N,F,\pathform \wedge (\smtvar[f']=\smtvar[f]\bitadd 64),\used\setminus\{\smtvar[f]\}),\smtvar[f] & \text{ if } \ty{ty} = \ptr{\tyi{n}} \\
    \end{cases} \and
    \semptradd[\domainsymbolic](\smtvar[b_1],k) = \smtvar[b_1] \bitadd k

  \end{mathpar}
  
  \caption{Evaluation and setting registers in symbolic context.}
  \label{fig:evalsetsymb}
\end{figure*}

\newcommand{\loadinner}[1]{\ensuremath{\mathtt{SymbLoad}^{#1}}}
\newcommand{\storeinner}[1]{\ensuremath{\mathtt{SymbStore}^{#1}}}

\begin{figure*}[t]
  \centering
  \begin{mathpar}
    \semload{\tyi{n}}{\domainsymbolic}((\smtvar[M],\smtvar[f],N,F,\pathform,\used),i)
    = \loadinner{\tyi{n}}(\smtvar[M],F(i)) \\ \and
    \loadinner{\tyi{n}}(\smtvar[M],\smtvar[a]) =
    \begin{cases}
      \arrselect (\smtvar[M],\smtvar[a])& \text{ if } \tyi{n} =
      \tyi{8}
      \\
      \arrselect (\smtvar[M],\smtvar[a])\bitconcat
      \loadinner{\tyi{n-8}}(\smtvar[M],\smtvar[a]\bitadd 1 ) & \text{
        otherwise}
    \end{cases}
  
    \semstore{\ty{ty}}{\domainsymbolic}((\smtvar[M],\smtvar[f],N,F,\pathform,\used),\smtvar[v],\smtvar[p])
    = (\smtvar[M'],\smtvar[f],N,F,\pathform\wedge (\smtvar[M'] =
    \storeinner{\ty{ty}} (\smtvar[M],\smtvar[v],\smtvar[p]))
    ,\used,\smtvar[v],\smtvar[p])

    \storeinner{\tyi{n}} ((\smtvar[M],\smtvar[v],\smtvar[p])) =
    \begin{cases}
      \arrstore(\smtvar[M],\smtvar[p],\smtvar[v]) & \text{ if }
      \tyi{n} =
      \tyi{8} \\
      \storeinner{\tyi{n-8}}
      (\arrstore(\smtvar[M],\smtvar[p],\smtvar[v][0\dots
      8]),\smtvar[p]\bitadd 1, \smtvar[v][8\dots n ])& \text{ otherwise
      }
    \end{cases}

  \end{mathpar}
  
  \caption{Store and Load operations in the symbolic context}
  \label{fig:loadstoresymb}
\end{figure*}

\begin{example}
  We briefly return to the module ($\module$) in \autoref{lst:symbneed} and consider how
  we can use the symbolic representation of \lodin\ to determine if
  the function \funcname{error} can be called. We simply
  instantiate the symbolic transition system $\lts{\domainsymbolic}{\module} =
(\netstates[\domainsymbolic],\netinitstate[\domainsymbolic],\systranssymb{})$
and generate symbolic states from $\netinitstate[\domainsymbolic]$ until we reach a state $\netstate[f] =
(s_1:s_2\dots:\emptylist,\domainstate[\domainsymbolic],\module)$ where $s_1 =
(\funcname{main},\lllabel{prev},\lllabel{call},\pc,\registermap,\frees)$
and $\domainstate[\domainsymbolic] = (\smtvar[M],\smtvar[f],N,F,
\pathform,\used)$.  Reaching $\netstate[f]$ reveals that there is a path in the control
flow graph of $\funcname{main}$ that reaches the call-block (and
thereby the call instruction), but not that it is feasible. To ensure
the feasibility, we invoke a SMT-solver and checks if
$\pathform$ is satisfiable. If this is the case, we can read the value
of all registers used along that path from the SMT satisyfing
assignment. 
\end{example}

\begin{remark}
  The symbolic context assigns each register of an \llvm\ program a
  single SMT-variable, and gathers constraints over these
  SMT-variables in a path formula. Assignments to \llvm\ registers is
  captured by equality between the SMT-variable and SMT-expressions. A
  result of this is that the symbolic context does not support
  assigning to the same register multiple times thus it  is only applicable for
  for programs without any loops in their
  control-flow-graph. 
\end{remark}

\paragraph{Merging Symbolic States}
\newcommand{\mergef}{\ensuremath{\mathtt{merge}}}
It is usual convenient to merge symbolic context states into one
state. This allows exploring several computational paths simultaneously and
helps combat path-explosion problem - which  is a big problem for
symbolic execution engines such as \klee.

For merging context-states \[\domainstate[\domainsymbolic] = 
(\smtvar[M],\smtvar[f],N,F, \pathform,\used)\] and
\[\domainstate[\domainsymbolic]' = (\smtvar[M]',\smtvar[f]',N',F',
\pathform',\used')\] where for all $n\in N\cap N'$ it is the case that $F(n) =
F'(n)$  we introduce the function $\mergef :
\domainstates[\domainsymbolic]\times
\domainstates[\domainsymbolic]\rightarrow
\domainstates[\domainsymbolic]$ defined as

\begin{align*}
\mergef(\domainstate[\domainsymbolic],&\domainstate[\domainsymbolic]')
= (\smtvar[M]'',\smtvar[f]'',N\cup N',F'',(\pathform\vee\pathform') \wedge
\pathform''\wedge \pathform''',\\
&\used\cup\used'\cup\{\smtvar[M]'',\smtvar[f]'',\smtvar[P]\})
\end{align*}
where  
\begin{itemize}
\item $F''(n) =
  \begin{dcases}
    F(n) & \text{ if } n\in N\\
    F'(n) & \text{ if } n\in N'\\
  \end{dcases}$
\item $\smtvar[M]'',\smtvar[f]'',\smtvar[P]\notin\used\cup\used'$,
\item $\syntaxeq{\pathform''}{\smtvar[M]'' = \smtite{\smtvar[P]}{\smtvar[M]}{\smtvar[M']}}$,
\item $\syntaxeq{\pathform'''}{\smtvar[f]'' = \smtite{\smtvar[P]}{\smtvar[f]}{\smtvar[f']}}$. 
\end{itemize}

Here $\smtvar[P]$ with type $\tyi{8}$ is a fresh SMT variable and
$\smtite{\smtvar[P]}{\smtvar}{\smtvar'}$ evaluates to $\smtvar'$ if
  $\smtvar[P] = \semfalse{\domainsymbolic}$ and to $\smtvar$ otherwise.
\section{Explicit Reachability Checking}
Model Checking~\citep{CGP99,DBLP:books/daglib/0020348} is a technique
widely used in  academia for validating that a formal model of a program
behaves correctly - according to a specification given by a logical
formula. A basic specification is a reachability specification, where
we are interested in finding a state where a given proposition
is true. This is the main focus in \lodin, and thus we will limit our
discussion to this setting.

\subsection{General Reachability Checking}
At the core of any reachability checking algorithm is a transition system to
search and a set of atomic propositions. In the case of \lodin, the state space we search is $\lts{\domainexplicit}{\module} = (\netstates,\netinitstate,\systransexpli{})$. Atomic propositions of a program are elements that may be true or
false in a state ( for instance whether $x==5$ or if a state has a
$\prop{DataRace}$ )\footnote{We define the exact propositions of \lodin\ in a short while}. An interpretation  (over states $\netstates$) of
an atomic proposition, $\prop{p}$,  is a function $\propfunc{\prop{p}}
: \netstates \rightarrow \{\truee,\falsee\}$, where $\truee$ indicates
$\prop{p}$ is true and $\falsee$ indicates it is false.  Atomic
propositions may be combined with the classical boolean operators
$\wedge,\vee$ and $\lnot$. The interpretation of these combined
propositions are defined recursively below as,
\begin{itemize}
\item $\propfunc{\psi_1 \wedge \psi_2} (\netstate) = \propfunc{\psi_1} (\netstate)
  \wedge \propfunc{\psi_2} (\netstate)$
\item $\propfunc{\psi_1 \vee \psi_2} (\netstate) = \propfunc{\psi_1} (\netstate)
  \vee \propfunc{\psi_2} (\netstate)$
\item  $\propfunc{\lnot \psi_1 } (\netstate) = \lnot\propfunc{\psi_1}
  (\state)$,   
\end{itemize}
where $\psi_1,\psi_2$ are  combined proposition themselves. 
\newcommand{\passed}{\ensuremath{\mathtt{Passed}}}
\newcommand{\waiting}{\ensuremath{\mathtt{Waiting}}}
Checking reachability for the proposition $\psi$  is now to check whether we, from the initial
state, can reach a state $\netstate$ where $\propfunc{\psi}(\netstate) =
\truee$. The classical approach for such a search is the fix-point
algorithm in \autoref{alg:reachFixP}.

For a finite state system
\autoref{alg:reachFixP} obviously terminate, as \passed\
eventually contains the entire reachable state space - and thus no  
further states can be put into \waiting\ and therefore $\waiting$ will
eventually become $\emptyset$. Equally straightforward is it to realise that
\autoref{alg:reachFixP}  produces correct results. 
\begin{algorithm}[tb]
  \KwData{Property : $\phi$}
  \KwData{Initial state: $\netstate$}
  \KwResult{$\top$ or $\bot$}
  $\passed := \emptyset$\;
  $\waiting := \{\netstate\}$\;
  \While{$\waiting\neq \emptyset$}{
    Let $\netstate[c]\in\waiting$\;
    $\waiting := \waiting\setminus\{\netstate[c]\}$\;
    \If{$\propfunc{\phi}(\netstate[c])$}{
      \Return{$\top$}
    }
    $\waiting := \waiting\cup\{\netstate \mid \exists i,\instr \text{s.t. } \netstate[c]\systransexpli[i]{\instr} \netstate \}$\;
    $\waiting := \waiting\setminus\passed$
  }
  \Return{$\bot$}
  \caption{The classic reachability algorithm. States that has not 
    been explored (but found) are kept in the set \waiting, and states that has
    already been processed are kept in \passed. 
\label{alg:reachFixP}}
\end{algorithm}
\autoref{alg:reachFixP} is non-deterministic in selecting an element
from $\waiting$ and in generating successors of the currently
considered state. The latter can easily be determinisied by generating states
in a fixed order, while the prior can be determinised in
different ways: the two usual ways is to keep the elements of \waiting\
in a stack or on a queue and let the order induced by these define the
search order.
\begin{remark}
  As mentioned earlier, the explicit state space may in fact be
  infinite thus  \autoref{alg:reachFixP} may not terminate. In \lodin\
  we have added options for terminating any verification after a user
  defined time or after using a user defined size of memory. 
\end{remark}

\paragraph{\llvm\ Propositions}
\lodin\ has support for propositions specifying classic programming errors
(division by zero, data race, out of bounds errors, etc). Furthermore,
it is posible to do comparisons between registers and check if a
specific function is called by a process. The use case for the latter
is, that the user can modify the verified program to call an
\texttt{error} function and check if that function is called
\footnote{This modification could even be  done at compile-time, by
  replacing the implementation of the commonly used \texttt{assert}
  function }.
\begin{figure}[tb]
  \centering
  {\scriptsize
  \begin{bnf*}
    \bnfprod{Prop}{ \bnfpn{Compare} \bnfor \bnfpn{Simple} }\\
    \bnfprod{Compare}{( \bnfpn{Comparand} \bnfpn{OP} \bnfpn{Comparand} )}\\
    \bnfprod{Comparand}{\bnfpn{Number} \bnfor \bnfpn{Register}}\\
    \bnfprod{Number}{\bnfpn{Integer};\bnfpn{Type}}\\
    \bnfprod{Register}{@\bnfpn{Integer}.\bnfpn{String}.\%\bnfpn{String};\bnfpn{Type}}\\
    \bnfprod{Type}{\bnfpn{us}8 \bnfor \bnfpn{us}16 \bnfor \bnfpn{us}32 \bnfor \bnfpn{us}64}\\
    \bnfprod{us}{ui \bnfor si}\\
    \bnfprod{OP}{< \bnfor <= \bnfor >= \bnfor > \bnfor == \bnfor !=} \\
    \bnfprod{Simple}{\datarace \bnfor \divzero \bnfor \overflows \bnfor [\bnfpn{Integer}.\bnfpn{String}]}
  \end{bnf*}
}
\caption{Grammar generating verification queries of \lodin.}

\label{fig:prorules}
\end{figure}
In \lodin s propositional language, registers and numbers are
typed to signed bitvectors or unsigned bitvectors with the suffixes \utyi{n} and \styi{n} where
$n\in\{8,16,32,64\}$. For any production rule $R$ in
\autoref{fig:prorules} we write $\lang{R}$ for the language generated
by that rule. An expression like
$\locregister{0}{F}{tmp3}{ui32} == \numberr{3}{ui32}$, means
\emph{take register \reg{\%tmp3} in the function \funcname{F} of the 0th
process. Interpret it as an unsigned 32bit integer, and compare it for
equality with 3 also interpreted as a unsigned 32bit integer}. For
comparisons to make sense, the two expressions being compared must,
naturally, have the same type.

For evaluating the value of a register
in  a state ($\netstate = (s_0,s_1,\dots,s_n,\domainstate,\module)$),
we define 
\begin{mathpar}
  \aritfunc{\locregister{k}{F}{tmp}{uin}} (\netstate ) =
  \begin{cases}
    \unsignedBV{\semevalreg{\domainexplicit}(\domainstate,r)} &\text{ if } \substack{s_k =
      ((\funcname{F},\prev,\cur,\pc,\registermap,\frees),\stackk) \\
      \reg{tmp}\typej \tyi{n}\\
      r = \registermap(\reg{tmp}) 
    }\\
    \zerovec{n} &\text{ otherwise}
  \end{cases}\\
  \aritfunc{\locregister{k}{F}{tmp}{sin}} (\netstate ) =
  \begin{cases}
    \signedBV{\semevalreg{\domainexplicit}(\domainstate,r)} &\text{ if } \substack{s_k =
      ((\funcname{F},\prev,\cur,\pc,\registermap,\frees),\stackk) \\
      \reg{tmp}\typej \tyi{n}\\
      r = \registermap(\reg{tmp}) 
    }\\
    \signedBV{\zerovec{n}} &\text{ otherwise}
  \end{cases}\\  
\end{mathpar}
Notice that we assign the default value of zero to
registers that are not present in the current activation
record. If the register is present in the activation record, we just
extract the bitvector and apply the interpretation function for
signed/unsigned numbers.

For evaluating numbers (e.g. \numberr{3}{ui32}) we write
$\aritfunc{\numberr{3}{ui32}}(\netstate)$ and it has the obvious
implementation. Given these notations, we can define how propositions
are evaluated within \lodin\ in \autoref{fig:prop_eval}.
\begin{figure}[tbp]
  \centering
  \begin{mathpar}
    \propfunc{A_1 \bowtie A_2} (\netstate ) =  \aritfunc{A_1} \bowtie
    \aritfunc{A_2}  \and
    \propfunc{\divzero} (\netstate) =
    \begin{dcases}
      \top &\text{ if }  \substack{\exists s_i =
        ((\lodinfunction,\prev,\cur,\pc,\registermap,\frees),\stackk)\\
        \lodinfunction =
        (\funcname{F},\registers,\parameters,\blocklabels,\blocksanon,\blockmap,\ty{ret}) \\
        \blockmap(\cur)[\pc] =
        \llbin{\%res}{DIV}{ty}{\%inp1}{\%inp2}\\
        \op{DIV} \in \{\op{udiv},\op{sdiv},\op{urem},\op{srem}\}\\
        r = \registermap(\reg{\%inp2})\\
        \unsignedBV{\semevalreg(\domainstate,r)} = 0 
      }\\
      \bot &\text{ otherwise} \\
    \end{dcases} \and
    \propfunc{\overflows} (\netstate) =
    \begin{dcases}
      \top &\text{ if }  \substack{\exists s_i =
        ((\lodinfunction,\prev,\cur,\pc,\registermap,\frees),\stackk)\\
        \lodinfunction =
        (\funcname{F},\registers,\parameters,\blocklabels,\blocksanon,\blockmap,\ty{ret}) \\
        \blockmap(\cur)[\pc] =
        \llstore[ty]{\%res}{\%inp1}{\%inp2}\\
        r = \registermap(\reg{\%inp2})\\
        \registermap(\reg{\%inp1}) \in \boolv^{l}\\
        (\texttt{len},v) = \memblockmap(\texttt{block}(r)) \\
        \texttt{offset}(r) + l > \texttt{len}
      }\\
      \top &\text{ if }  \substack{\exists s_i =
        ((\lodinfunction,\prev,\cur,\pc,\registermap,\frees),\stackk)\\
        \lodinfunction =
        (\funcname{F},\registers,\parameters,\blocklabels,\blocksanon,\blockmap,\ty{ret}) \\
        \blockmap(\cur)[\pc] =
        \llstore[ty]{\%res}{\%inp1}{\%inp2}\\
        r = \registermap(\reg{\%inp2})\\
        \registermap(\reg{\%inp1}) \in \boolv^{l}\\
        \bot = \memblockmap(\texttt{block}(r)) \\

      }\\
      
      \bot &\text{ otherwise} \\
    \end{dcases}\and

    \propfunc{[i.func]} (\netstate) \begin{dcases}
      \top &\text{ if }  \substack{\exists s_i =
        ((\lodinfunction,\prev,\cur,\pc,\registermap,\frees),\stackk)\\
        \lodinfunction = (\funcname{F},\registers,\parameters,\blocklabels,\blocksanon,\blockmap,\ty{ret}) \\
        \blockmap(\cur)[\pc] = \reg{\%res}\,=\, \op{call}\, \ty{ret}\, \reg{@func}\, (\ty{ty1}\, \reg{\%\hat p1} \dots \ty{tyn}\, \reg{\%\hat pn})\\
      }\\
      \bot &\text{ otherwise} \\
    \end{dcases} \and
    
    \propfunc{\datarace} (\netstate) =
    \begin{dcases}
      \top &\text{ if }  \substack{\exists s_i =
        ((\lodinfunction[i],\prev[i],\cur[i],\pc[i],\registermap[i],\frees[i]),\stackk_i)\\
        \lodinfunction[i] =
        (\funcname{F}_i,\registers[i],\parameters[i],\blocklabels[i],\blocksanon[i],\blockmap[i],\ty{ret_i}) \\
        \blockmap[i](\cur)[\pc[i]] = \llload[ty_1]{res}{ptr_i}
        \\
        p_i =  \semevalreg[\ptr{ty_i}]{\domainexplicit} (\domainstate,\registermap[i](\reg{ptr_i}))\\
        \exists s_j = ((\lodinfunction[j],\prev[j],\cur[j],\pc[j],\registermap[j],\frees[j]),\stackk_j)\\
        \lodinfunction[j] =
        (\funcname{F}_j,\registers[j],\parameters[j],\blocklabels[j],\blocksanon[j],\blockmap[j],\ty{ret_j}) \\
        \blockmap[j](\cur)[\pc[j]] = \llstore[ty_j]{res}{val}{ptr_j}
        \\
        p_j =  \semevalreg[\ptr{ty_j}]{\domainexplicit} (\domainstate,\registermap[i](\reg{ptr_i}))\\
        \texttt{block}(p_i) = \texttt{block}(p_j) \\
        \{\texttt{offset}(p_i), \dots \texttt{offset}(p_i)+\bytesize(\ty{ty_i}) \} \cap \\ \{\texttt{offset}(p_j), \dots \texttt{offset}(p_j)+\bytesize(\ty{ty_ij}) \} \neq \emptyset
      }\\
      \bot &\text{ otherwise} \\
    \end{dcases}
  \end{mathpar}
  \caption{Evaluation of propositions in \lodin\ where
    $A_1,A_2\in\lang{Register}\cup\lang{Number} ,\bowtie\in\lang{OP}$,
    $\netstate = (s_0,s_1,\dots,s_n,\domainstate,\module)$ and \domainstate = ((\memblockmap,\memused),N,F)  \label{fig:prop_eval}. For $\overflows$ we have only shown the rule for overflows at writes, but naturally there is an equivalent rule for reads. }
\end{figure}
A short discussion may be in order about the evaluations in
\autoref{fig:prop_eval}.
\begin{itemize}
\item Division by zero (\divzero) are determined in the
obvious manner, where we simply check if any process executes any
instruction involving a
division\footnote{\op{div},\op{sdiv},\op{rem},\op{srem}} and check if
the second operand is zero.
\item Buffer overflows (\overflows) are
likewise easily checked by checking if any process accesses memory,
and for each of those that do access memory we check if their
read/write to memory exceeds the length of the buffer they are
writing/reading into/from.
\item The instruction for checking whether a specific process number $i$
can call a function \funcstyle{func} $([i.\funcstyle{func}])$, we
first check if process $i$ performs a \op{call} instruction and if so,
if the functions being called matches $\funcstyle{func}$.
\item The most diffuclt proposition to check is without a doubt \datarace. For
evaluating this instruction, we iterate over all processes and finds
pairs of read/write and  write/write to the samme pointer
base. Afterwards we check if their $\mathtt{offset}+\mathtt{length}$
overlaps
\end{itemize}
 
\begin{example}
  As a short example of using \lodin\ for reachability checking let us
  consider  \autoref{lst:example} and consider we are interested in
  whether $\reg{\%x}$ and $\reg{\%z}$ can ever be equal. Notice that
  since   all
  \op{phi} instructions should be  executed atomically in the
  beginning of a  block, this should never be possible - thus checking
  this with \lodin\  actually checks if \lodin\ implements the \op{phi}
  instructions behaviour correctly.

  In \lodin\ we can check the property by asking the query
  $\mathtt{E<>}
  (\locregister{0}{main}{x}{ui32} == \locregister{0}{main}{z}{ui32})$. 

Unfortunately \lodin\  reports that this is indeed possible
even though it should not be. There is a logical explanation for this:
both registers are initialised by  \lodin\ to $0$ thus in the initial state
they are equal. For this reason, it is more reasonable to use the
$\reg{\%b}$ register for our check thus we check the query
$\mathtt{E<>} (\locregister{0}{main}{b}{ui8} == \numberr{1}{ui8})$ and
get the result in \autoref{lodterm:example2} indicating it is indeed not possible. 
\begin{lstterm}
\begin{cmd}
\begin{term}
$Lodin example.ll example2.q
Lodin 0.3 (Jul  8 2019)
Revision: 0.2-802-ga42644cf
Importance Ratio: double
LLVM: 8.0.0

LLVM module modifications: 
Remove Unuused instructions
Warning: No entry-point specified. Assuming main.
Random seed: 1562587068
System: NaiveGraph-explicit
Platform: PThread
Storage: SharedMem Storage
Successor: Standard
Prob-Successor: Standard
Passed-Waiting: Standard
SMT-Backend: Boolector 3.0.0

Verifying: E<>((0.main.b==) )
Warning: Casting register main.b to integer type UI8 - can't guarantee LLVM uses this register as such

Not Satisfied
\end{term}
\end{cmd}
\caption{Output from \lodin.}\label{lodterm:example2}
\end{lstterm}
\end{example}

\subsection{State Space Reductions}
A well-known problem for explicit-state reachability checking of
parallel systems is the notorious \emph{state space explosion problem}
i.e. that the combined state space increases exponentially when each
process of the system increases linearly. This is a huge problem when
considering high-level programs and exacterbated when using \llvm\ as
input, because \llvm\ programs has more instructions per process. For
making explicit-state reachability checking possible we thus need ways
of limiting the size of the state space. A first realisation to reduce
the state space is, that processes can only influence each others
behaviour at predefined points, namely when accessing
memory.  Due to our specification language
allowing to query whether functions can be called, we also consider
\op{call} instructions  to affect the external behaviour of a
process. We say that an instruction $\instr$  is internal if
$\instr$ if it is a \op{load},\op{store} or $\op{call}$ instruction.  We denote
the set of all internal instructions by
$\internalinst{\registers}$. In the following we describe the two
state space reductions that are implemented inside \lodin. They both
define a new transition relation, that can directly replace
$\systransexpli{}$. 

\begin{description}
\item[\tausymb] Our first state space reduction is based on the idea,
  that when a process performs a transition step it will perform all
  following transitions that executes internal instructions. More
  formally, we replace the transitions relation $\systransexpli{}$
  with $\systranstau{}$ where $\systranstau{}$ is defined according to the
  rule
  \[\transrule{\forall k>2,
    \instr[k]\in\internalinst{\registers}}{\left[ \netstate[k-1]
      \systransexpli[i]{\instr[k]} \netstate[k] \right]_{k=1\dots
      n}}{\netstate[0] \systranstau[i]{\instr[1]\dots\instr[n]}
    \netstate[n]}.\]

  Notice, that there is no lower length in then size of the sequence
  $\instr[1],\dotsm\instr[n]$. To achieve the largest reduction,
  \lodin\ always uses the longest  possible sequence.
  
\item[\ntausymb] In this state space reduction, all processes that perform
  internal instructions execute simultaneously while all other
  processes execute independently. The transition relation
  $\systransntau{}$ is defined by two rules
  \[
    \transrule{\forall k,
    \instr[k]\in\internalinst{\registers}}{\left[ \netstate[k-1]
      \systransexpli[i_k]{\instr[k]} \netstate[k] \right]_{k=1\dots
      n}}{\netstate[0] \systransntau[i_1,\dots,i_n]{\instr[1]\dots\instr[n]}
    \netstate[n]}\] 

\[\transrule{\instr\notin\internalinst{\registers}}{\netstate
      \systrans[i]{\instr} \netstate'}{\netstate \systransntau[i]{\instr}
    \netstate'}\]
\end{description}

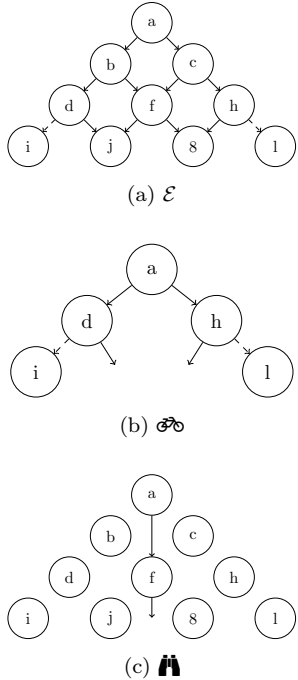
\begin{figure}[tb]
  \centering
  \subfloat[\standardsymb]{
  \resizebox{.25\textwidth}{!}{
  \begin{tikzpicture}
    \matrix[ampersand replacement=\&]
    {
      \&\&  \& \node[state] (a){a}; \&  \& \& \\
      \&\& \node[state] (b) {b}; \& \& \node[state] (c) {c}; \& \&  \\
      \& \node[state] (d) {d}; \&  \& \node[state] (f) {f}; \&   \& \node[state] (h) {h}; \& \\   
      \node[state] (i) {i};\& \& \node[state] (j) {j};   \&  \& \node[state] (k) {8};    \& \& \node[state] (l) {l};;  \\
    };
    
    \draw[->] (a) edge (b);
    \draw[->] (b) edge (d);
    \draw[->,dashed] (d) edge (i);
    
    \draw[->] (a) edge (c);
    \draw[->] (c) edge (h);
    \draw[->,dashed] (h) edge (l);
    
    \draw[->] (b) edge (f);
    \draw[->] (c) edge (f);
    \draw[->] (f) edge (j);
    \draw[->] (f) edge (k);
    \draw[->] (d) edge (j);
    \draw[->] (h) edge (k);
  \end{tikzpicture}
}
}

\subfloat[\tausymb]{
\resizebox{.25\textwidth}{!}{
      \begin{tikzpicture}
        \matrix[ampersand replacement=\&]
        {
          \&\&  \& \node[state] (a){a}; \&  \& \& \\
          \&\&  \& \&  \& \&  \\
          \& \node[state] (d) {d}; \&  \&  \&   \& \node[state] (h) {h}; \& \\   
          \node[state] (i) {i};\& \& \node (j) {};   \&  \& \node (k) {};    \& \& \node[state] (l) {l};;  \\
        };
        
        \draw[->,double] (a) edge (d);
        \draw[->,dashed] (d) edge (i);
        
        \draw[->,double] (a) edge (h);
        \draw[->,dashed] (h) edge (l);
        
        \draw[->] (d) edge (j);
        \draw[->] (h) edge (k);
      \end{tikzpicture}
    }
  }
  
\subfloat[\ntausymb]{
  \resizebox{.25\textwidth}{!}{
    \begin{tikzpicture}
      \matrix[ampersand replacement=\&]
      {
        \&\&  \& \node[state] (a){a}; \&  \& \& \\
        \&\& \node[state] (b) {b}; \& \& \node[state] (c) {c}; \& \&  \\
        \& \node[state] (d) {d}; \&  \& \node[state] (f) {f}; \&   \& \node[state] (h) {h}; \& \\   
        \node[state] (i) {i};\& \& \node[state] (j) {j};   \&  \& \node[state] (k) {8};    \& \& \node[state] (l) {l};;  \\
      };
      \node[below of = f] (bb) {};
      \draw[->] (a) edge (f);
      \draw[->] (f) edge (bb);
    \end{tikzpicture}
  }
}
\caption{\lodin\ state space reductions. Transitions going left
  originates from one process while transitions going to the right
  correspond to another. Dashed arrows indicate visible actions.}
\label{fig:reductions}
\end{figure}

In \autoref{fig:reductions} we provide a graphical overview of how
these reductions modifies the state space.

\begin{example}
  As an example of the state space reductions that \ntausymb\ and \tausymb\
  respectively do, consider the C-program in \autoref{fig:petersons} that executes
  petersons mutual exclusion algorithm. To use this program with
  \lodin, it must first be compiled to an \llf -file using \texttt{clang}
 \footnote{\texttt{clang -S -c -emit-llvm file.c}}. After this step we
 can inspect the state space reductions achieved by asking \lodin\ the query
 \texttt{EnumStates} on the resulting \llf -file with  the different
 state space reductions. In \autoref{tab:spacereductions} we see the
 reported number of states, along with how many states with data races
 that was encountered. Notice that \ntausymb\ in this case achieves the
 largest reduction. 
 \begin{table}[tb]
   \centering
   \begin{tabular}{l r r}
     \toprule 
     State Generator  &  States & DataRace States\\ \midrule
     \standardsymb & 6573 & 16 \\
     \tausymb & 4111 & 16 \\
     \ntausymb & 3057 & 8 \\
     \bottomrule
   \end{tabular}
   \caption{State Space Reductions.}
   \label{tab:spacereductions}
 \end{table}
\end{example}

\begin{figure*}[tb]
\begin{multicols}{2}
\begin{cprgm}
#include <stdio.h>

int flags[2] = {0,0};
int turn = 0;

void crit () {}

typedef struct {
  int *mflag;
  int *oflag;
  int* turn;
}Options;

void* petersons1 () {  
  Options opt;
  opt.mflag = &flags[0];
  opt.oflag = &flags[1];
  opt.turn = &turn;
      
  *(opt.mflag) = 1;
  *(opt.turn) = 1;
      
  while (*(opt.oflag)  && *(opt. turn) == 1)
	{
	  // busy wait
	}
  // critical section
  crit ();
  // end of critical section
      
  *(opt.mflag) = 0;
      
  return 0;
}

void* petersons2 () {
  Options opt;
  opt.mflag = &flags[1];
  opt.oflag = &flags[0];
  opt.turn = &turn;
       
  *(opt.mflag) = 1;
  *(opt.turn) = 0;
       
  while (*(opt.oflag)  && *(opt. turn) == 0)
	{
	  // busy wait
	}
  // critical section
  crit ();
  // end of critical section
       
  *(opt.mflag) = 0;
       
  return 0;
}
\end{cprgm}
\end{multicols}

\caption{Petersons Mutual Exclusion Protocol}\label{fig:petersons}
\end{figure*}

Although the above state space reductions can reduce the state space
due to interleavings dramatically, they cannot reduce the number of
states caused by non-deterministic input. A program with just one
non-deterministic 32bit value will end up having over $2^{32}$ states.


\section{Simulation-Based Model Checking}
In the preceding section we saw how \lodin\ can be used to perform an
exhaustive state space search under an explicit context. We also realised, that the state space explosion problem poses a problem
for any exhaustive search and showed how \lodin\ can reduce this
explosion through state space reductions. The state space reductinos
also have their limits  thus we need other strategies for handling
this explosion. \lodin\ proposes to use a simulation-based technique,
where random (step-bounded) traces are drawn from the program and
inspected for satisfaction of the property at hand. At the heart of
any simulation-based technique is an underlying simulation  
distribution. The simulation distribution may stem from actual knowledge of how the
system behaves, in which case simulations can be used to calculate
actual probabilities of the system satisfying the property using
statistical methods - hence the name statistical model checking~\cite{YKNP06}. In
case the simulation distribution is ``arbitrary'', then estimated
probabilities  are meaningless for the system itself, but serves as a
way to predict how likely it is that a continued search will find the
property searched for. In this case the technique is called Monte
Carlo Model Checking. 

\subsection{Simulation Distribution}
In \lodin\ each state $\netstate$ of the state space $\lts{\domainexplicit}{\module} = (\netstates,\netinitstate,\systransexpli{})$ is
assigned a probability distribution $\selectp{\netstate}
: \mathbb{N} \rightarrow [0,1]$.  The probability distribution assigns
a probability to
which process should perform an action. The function 
$\selectp{\netstate}$  should obviously only assign a
probability mass to a process if that process can perform a transition
thus we require that $\selectp{\netstate}(i) \neq 0 \implies \netstate
\systransexpli[i]{\instr} \netstate'$, for some $\netstate'$. Having
selected who should perform an action, we also need a probability
function for the result of that choice $i$. We do this by assuming a
$\selectv{\netstate,i} : \netstates \rightarrow [0,1]$, where
$\netstates$ is the set of all states. The requirement to this
function is, that it should only assign probabilities to states that
can be reached by the $ith$ process performing a transition from
$\netstate$ i.e. $\selectv{\netstate,i}(\netstate') \neq 0
\implies \netstate \systransexpli[i]{\instr} \netstate$ for some
instruction $\instr$.

Given these two probability mass functions, the probability that a
system generates the finite transition sequence $\omega =
\netstate[0]\systransexpli[i_1]{\instr[1]}\netstate[1]\systransexpli[i_2]{\instr[2]}\dots
\systransexpli[i_{n1}]{\instr[n]} \netstate[n]$, where $\netstate[0]$ is
the initial state, is given  by  $\prob(\omega) =
\prod_{k=1}^{n}\selectp{\netstate[k-1]}(i_{k})\cdot\selectv{\netstate[k-1],i_{k}}(\netstate[k])$. For
a transitions sequence $\omega  =
\netstate[0]\systransexpli[i_1]{\instr[1]}\netstate[1]\systransexpli[i_2]{\instr[2]}\dots
\systransexpli[i_{n1}]{\instr[n]} \netstate[n]$, we let $|\omega| = n$ be its
length and $\omega[i] = \netstate[i]$ . We also let $\Omega^{m,\module}$ be the set of all transition sequences $\omega$
with $|\omega| = m$ of \llvm\ module \module. Let $\prop{p}$ be a
proposition, and $\omega\in\Omega^{m,\module}$ then we define  the indicator function 
\[\mathbb{I}_{\prop{p}} (\omega) =
  \begin{cases}
    1 &\text { if } \exists $i$ \text{ s.t. }
    \propfunc{\prop{p}}(\omega[i]) = \truee \\
    0 &\text { otherwise}
  \end{cases}
\]
that returns $1$ if $\omega$ at some point satisfies $\prop{p}$ and 0
otherwise. With this at our hand, we define the probability that an
execution trace of a program $\module$ satisfies a proposition
$\prop{p}$ within $m$ steps as 

\[
  \mathtt{Pr}_{\module,m}(\prop{p}) = \sum_{\omega\in\Omega^{m,\module}}
  \mathbb{I}_{\prop{p}}(\omega)\cdot \prob(\omega)
\]

As the probability only depends on the state, we usually project out
transitions and only generate the states. An algorithms for
generating a sequence of states from $\netstate[0]$ according to the
probability distribution can be seen in \autoref{alg:sim}. In the
algorithm we use $k  \sim P$ to mean that $k$ is distributed according
to the probability mass function $P$. 
\begin{algorithm}[tb]
  \KwData{Initial state: $\netstate[0]$}
  \KwData{Length: $n$}
  $\omega = \netstate[0]$\;
  \For{$i\in\{1,\dots, n\}$}{
    $k \sim \selectv{\netstate[i-1]}$\;
    $\netstate[i] \sim \selectp{\netstate[i-1],k}(\netstate[i-1])$\;
    $\omega = \omega\netstate[i]$\;
 }
 \Return{$\omega$}
 \caption{Generating random traces in \lodin\label{alg:sim}}
\end{algorithm}

\begin{example}
  Before dwelling upon how to using simulation to do verification, let
  us briefly consider what kind of coverage of the state space we can
  expect with by doing simulations. To this end, we have implemented
  the query \texttt{EnumStatesSMC <=l n}. This query simply generates
  $n$ traces each of length $l$ and keeps tracks of how many different
  states it has visited in total. We show the results of running this
  query on \autoref{fig:petersons} in \autoref{tab:simulationStates}. Recall from previously, that the total number of states is $6573$.
  \begin{table}[tb]
   \centering
   \begin{tabular}{l r r}
     \toprule 
     $n$  &  States & DataRace States\\ \midrule
     1 & 77 & 1 \\
     100 & 1840 & 4 \\
     1000 & 3579 & 11 \\
     10000 & 4714 & 14 \\
     \bottomrule
   \end{tabular}
   
   \caption{State encountered with SMC. The used query is \texttt{EnumStatesSMC <=5000 $n$}. }
   \label{tab:simulationStates}
 \end{table}
  
\end{example}
\subsection{Statistical Model Checking}
Statistical model checking tries answering two questions:
\begin{inparaenum}
\item a quantitative ``What is the probability $\theta$ of reaching \prop{p}''?, and
\item a qualitative ``Is the probability of $\theta$ greater than $\theta_t$''? 
\end{inparaenum}
Both questions are answered by generating a number samples and using
statistical techniques to infer the answer with a user specified
confidence.

\paragraph{Quantitative}
Here we  repeatedly generate runs and construct 
an interval $[\theta_l,\theta_u]$ for which we are confident that the
probabiltity $\theta$
is contained within. For the following we assume we are provided
with $\epsilon$ being the wanted width of the
interval and an $\alpha\in[0,1]$ indicating the confidence $(1-\alpha)$ we want in
the interval.
\begin{lstterm}
\begin{cmd}
\begin{term}
Lodin 0.3 (Jul  8 2019)
Revision: 0.2-802-ga42644cf
Importance Ratio: double
LLVM: 8.0.0

LLVM module modifications: 
Remove Unuused instructions
Warning: Function signature of entry point petersons1 (Pointer()) does not match on return type by platform (UI32())
Warning: Function signature of entry point petersons2 (Pointer()) does not match on return type by platform (UI32())
Random seed: 1562589004
System: NaiveGraph-explicit
Platform: PThread
Storage: SharedMem Storage
Successor: Standard
Prob-Successor: Standard
Passed-Waiting: Standard
SMT-Backend: Boolector 3.0.0

Verifying: Pr[<=5000](<>DataRace )

Result: [0.285738,0.295738 ] with confidence 0.95
Total Runs: 31883, Satisfying Runs: 9269

Histogram: Satisfying Runs
Max Frequency:  0.504262
Values in [28, 103 ] in steps of 1
[ 4674, 2057, 0, 0, 0, 0, 0, 0, 0, 480, 0, 0, 0, 0, 0, 0, 0, 615, 0, 0, 0, 0, 84, 0, 0, 0, 0, 0, 0, 0, 20, 0, 0, 0, 0, 1167, 0, 0, 0, 0, 0, 0, 0, 0, 0, 0, 0, 0, 152, 0, 0, 0, 0, 0, 0, 0, 0, 0, 0, 0, 0, 18, 0, 0, 0, 0, 0, 0, 0, 0, 0, 0, 0, 0, 2 ]
\end{term}
\end{cmd}
\caption{\lodin-output}\label{lst:exampledata}
\end{lstterm}

Consider that we have generated a sequence of samples
$\omega_1,\omega_2,\dots,$ and let $x_1,\dots,x_m$ be random
variables
such that $x_i = \mathbb{I}_{\prop{p}}(\omega_i)$. Then each variable $x_i$ has a Bernoulli
distribution with success probability $\theta_t$ and the sum $X_m = \sum_{i=1}^m
x_i$ is binomially distributed. We construct a confidence interval
using the exact confidence interval by \citet{Clopper1934}: if we have
$m$ samples then a Clopper-Pearson-interval with confidence $\alpha$
is given as the intersection  $S_\leq\cap S_\geq$ where
\[
  S_\leq = \{\psi \mid \mathtt{B}_{m,\psi}(X_m) > \alpha/2 \}
\]
\[
  S_\geq = \{\psi \mid 1 - \mathtt{B}_{m,\psi}(X_m) > \alpha/2 \}
\]

\noindent and $\mathtt{B}_{m,\psi}$ is the cumulative distribution function for a binomial
distribution with $m$ samples and success parameter $\psi$. Notice
that we are not in control of the resulting width of this interval -
more samples will however shrink the width $\epsilon$ and thus we simply
iteratively produce samples until we get the desired
width.

\begin{example}
Let us consider the program in \autoref{fig:petersons} again and let
us asses the probability that a data race is encountered. We can asses
this with the query: \texttt{Pr[<=5000] (<> DataRace)}. The $5000$ in
this query is the length of the runs. See \autoref{lst:exampledata} for the output.
From the output we can see that \lodin\ estimates the probability to
lie in the interval $[0.29,0.30]$. The last part provides a histogram
over the length of the satisyfing runs.  \lodin\ runs by default with
$\alpha = 0.05$ and $\delta = 0.01$. These parameters can be tweaked
by suffixing the query with \texttt{\{Alpha = Float, Epsilon = Float\}}
where \texttt{Float} are numbers in $[0,1]$.  Running the query
\texttt{Pr[<=5000] (<> DataRace) \{Alpha = 0.01, Epsilon = 0.05\}} for
instance gives the result $ [0.27,0.32 ]$.
\end{example}

\paragraph{Qualitative.}
Checking whether the probability $\mathtt{Pr}_{\module,m}(\prop{p})$ exceeds a threshold $\theta$ 
can be answered by doing hypothesis testing. We test the hypothesis
$H_0: \mathtt{Pr}_{\module,m}(\prop{p})  \geq \theta$ against $H_1:\mathtt{Pr}_{\module,m}(\prop{p}) < \theta$. In advance, we want to define two parameters,
$\alpha$ (significance level) and $\beta$ (power level), that signifies how willing we are to reject a true
hypothesis and how willing we are to accept a false hypothesis. In practice
 we want a test for which the probability of rejecting
 $H_0$ while $H_0$ is true is less than $\alpha$; while the probability
 of accepting $H_0$ while $H_1$ is true is less than $\beta$.
\begin{algorithm}[t]
  \KwData{Initial State: $\state$}
  \KwData{Property: $\mathtt{Pr}_{\module,m}(\prop{p}) \geq \theta$}
  \KwData{Indifference Region: $2\cdot\delta$}
  \KwData{Significance Level: $\alpha$}
  \KwData{Power Level: $\beta$}
  \KwResult{$\top$ or $\bot$ } 
  $p_0 = \theta +\delta $\;
  $p_1 = \theta - \delta$\;
  $r = 0$\;
  \While {$ d > \delta$}{
   
    $\omega = generateRun (\state,m)$\;
    $x =  \mathbb{I}_{\prop{p}}(\omega)$\;
    $r = r+x\cdot log(P_1/p_0)+(1-x)\cdot log((1-p1)/(1-p_0))$\;
    \If{$r\leq log (\beta/(1-\alpha))$}{
      \Return{$\top$}
    }
    \If{$r\geq log((1-\beta)/\alpha)$}{
      \Return{$\bot$}
    }
    
  }  
  \caption{Testing whether probability is larger than $\theta$ \label{agl:hypo}}
\end{algorithm}
 Realising that acheiving both of these requirements is close to
 impossible in general~\citep{younesThesis} we introduce an
 indifference region of width $2\cdot \delta$ around $\theta$ and test
 instead the hypothesis $H_0' \mathtt{Pr}_{\module}(\prop{p}) :  \phi \geq \theta + \delta$ against $H_1' :  \mathtt{Pr}_{\module}(\prop{p}) < \theta - \delta$. \citet{Wald1973} developed a sequential
hypothesis testing algorithm, see \autoref{agl:hypo},  for exactly this case; the idea is to
iteratively generate runs and based on these calculate a value $r$ -
 eventually this value will cross $log (\beta/(1-\alpha))$ or
$log((1-\beta)/\alpha)$ and $H_o'$ is either rejected or accepted.


\section{Bounded Model Checking}
In previous sections we described the symbolic representation of
states used within \lodin, and we saw in an example how this
representation could be used to explore many values registers
simultaneously. We however did not give a structured way of using this
symbolic representation in a verification framework. We make up
for that in this section. 
\newcommand{\mergees}{\ensuremath{\mathtt{Mergees}}}
\subsection{Symbolic Analysis of Loop-free program}
\begin{algorithm}[htbp]
  \KwData{Property : $\phi$}
  \KwData{Initial state: $\netstate$}
  \KwResult{$\top$ or $\bot$}
  $\mergees := \mergees$\;
  $\waiting := \{\state\}$\;
  \While{$\waiting\neq \emptyset$}{
    Let $\netstate[c]\in\waiting$\;
    $\waiting := \waiting\setminus\{\netstate[c]\}$\;
    \If{$\propfunc{[i.\funcname{func}]}(\netstate[c])$}{
      \Return{$\top$}
    }
    \ForEach{$\netstate[n] \in \{\netstate \mid \exists i,\instr
      \text{s.t. } \netstate[c]\systranssymb[i]{\instr} \netstate\}$}{
      \If{$\lnot\mathtt{Mergeable}(\netstate[n])$} {
        $\waiting = \waiting\cup\{\netstate[n]\}$\;
      }
      \Else {
        Let $\netstate[n] =
        ((\lodinfunction,\prev,\cur,\pc,\registermap,\frees):S,\domainstate[\domainsymbolic])$
        \;
        \If {$\exists (\cur,\netstate[o],n)\in\mergees$}{
          \If{$n-1 = 0$}{
            $\waiting = \waiting\cup\{\mergef(\netstate[o],\netstate[n])\}$\;
          }
          \Else {
            $\mergees =
            \mergees\setminus\{(\cur,\netstate[o],n))\}\cup\{
            (\cur,\mergef(\netstate[o],\netstate[n]),n-1)$
            \}
          }
        }
        \Else {
          $\mergees =  \mergees\cup\{(\cur,\netstate[n],\incoming(\netstate[n])-1)\}$\;
        }
      }
    }

  }
  \Return{$\bot$}
  \caption{The symbolic reachability algorithm. \label{alg:reachFixP}}
\end{algorithm}
In this section we show how \lodin\ uses its symbolic representation to analyse
single-threaded programs without loops. For now, we will also restrict
our attention to verify if a given function can be called at any
time \eg propositions as $[0.\funcname{error}]$. Before going into
details about the algorithm, we will setup up some convenient
notations, to make the algorithm more readable. 

A key concept we will need in the algorithm for analysing loop-free
programs is converging basic blocks and diverging basic blocks: for a
\llvm\ function
$(\funcname{N},\registers,\parameters,\blocklabels,\blocksanon,\blockmap,\ty{ret})$,
we say that a block $\block\in\blocksanon$ diverges control flow if
$\syntaxeq{\block[|\block|]}{\llbrcond{c}{trueb}{falseb}}$. For a block
$\block\in\blocksanon$ where $\blockmap(\lllabel{con}) = \block$ for
some $\lllabel{con}$, we define the set of all blocks jumps to $\block$ as  

\begin{align*}
  \incoming(\lllabel{con}) = &\{ \block'\in\blocksanon \mid
  \syntaxeq{\block'[|\block'|]}{\llbrcond{r}{con}{f}}  \}\cup \\
  & \{ \block'\in\blocksanon \mid
    \syntaxeq{\block'[|\block'|]}{\llbrcond{r}{t}{con}}  \} \cup \\
  &\{ \block'\in\blocksanon \mid
  \syntaxeq{\block'[|\block'|]}{\llbr{con}}  \},
\end{align*}

and say that $\lllabel{con}$ labels a converging block if
$|\incoming(\lllabel{con})| > 1$. For ease of writing we will say that
$\lllabel{con}$ is a converging block. The definition of $\incoming$
we lift to states of $\lts{\domainsymbolic}{\module} =
(\netstates[\domainsymbolic],\netinitstate[\domainsymbolic],\systranssymb{})$
as follows: if $\netstate = (s_1:S,\domainstate[\domainsymbolic])$ and
$ s_1 = (\lodinfunction,\prev,\cur,\pc,\registermap,\frees)$ then
$\incoming(\netstate) = \incoming(\cur)$.

In the discussion of the symbolic context, we defined how to merge
symbolic context states. Here we wish to lift merging to a state
$\netstate, \netstate' \in \netstates[\domainsymbolic]$. A state
$\netstate =
((\lodinfunction,\prev,\cur,\pc,\registermap,\frees):S,\domainstate[\domainsymbolic])$
is considered \emph{mergeable} (written $\mathtt{Mergeable(\netstate)}$) if $\pc$ is not a \llphiname\
instruction and $\incoming(\netstate)> 1$. It can be merged with another state $\netstate' =
((\lodinfunction,\prev',\cur,\pc,\registermap,\frees'):S,\domainstate[\domainsymbolic]')$
if $\domainstate[\domainsymbolic]$ and
$\domainstate[\domainsymbolic]'$ can be merged. The $merge$ of
$\netstate,\netstate'$ is defined as:
\begin{align*}
  \mergef (\netstate,\netstate') =&
  ((\lodinfunction,\prev,\cur,\pc,\registermap,\frees\cup\{\frees'\}):S,\\
  & \mergef
  (\domainstate[\domainsymbolic],\domainstate[\domainsymbolic]'))
\end{align*}

After these preliminary setups, we are ready to show the algorithm in
\autoref{alg:reachFixP}. To a large extend it is the classic
reachability algorithm where unexplored states are kept in a \waiting
list, and immediately after being pulled from the \waiting, is is
checked if the property at hand is satisfied. Checking if the property $[i.\funcname{func}]$
is true involves
\begin{enumerate}
\item checking if the function \funcname{func} is being called by the
  ith process (a
  check that does not depend on the \llvm registers),
\item checking if the path formula of the state is satisfiable. 
\end{enumerate}

If the property is not satisfied, then all possible successor are
generated and either put into \waiting (if  not a
$\mathtt{Mergeable}$ state) or it is tried merged with a state already in a \mergees\ queue.  

\paragraph{Handling Loops}
Any nontrivial program will have loops, and as such verificaion
techniques must cope with loops. \lodin\ can verify programs with loops, but
relies on syntactially unrolling the loops before verification. In case
the loop unroll is complete, then the verification is complete -
otherwise the verification is only sound.

\section{Implementation Details}
\lodin - available at \downloadurl\  -  is build around the \llvm -bitcode and uses the \llvm
-libraries for parsing  the input-files, and performing some \llvm\ modifications during. \lodin\ does, however, 
not use the infrastructure of \llvm\ for performing analyses.
Instead it builds its own internal representation of the
loaded \llvm module and implements its own state space successor
generator.

\subsection{\llvm\ Modifications}
At load time \lodin\ can perform a number of modifications of the
\llvm\ program - some of the modifications are enabled by default,
some forced enabled by others\footnote{To help the user, the modified program
  can be outputted at load time as well}. In the following we briefly discuss the
modifications.

\paragraph{Naming Instructions}
\llvm -bitcode files do not necessarily contain names for the
registers. At load time \lodin\ therefore give names to all non-named
registers in the program. This simplifies internally when providing
error messages.  

\paragraph{Constant Removal}
\llvm -bitcode instructions can have constant expressions  which the
interpreter of \lodin\ would have to evaluate at run time. We replace
these constant expressions with \llvm\ instructions thus simplifying
the subset of \llvm\ that our interpreter needs to understand. 

\paragraph{Simplify CFG}
This is a standard \llvm\ modification that attempts to simplify the
control flow graph. \lodin\ provides an option for running this
simplification, but does not run it by default as it modifies the
program drastically and thus specifications of the user is perhaps no
longer ``valid''. The modification can be enabled by the user or
\emph{forced} by other modifications.

\paragraph{Elimninate Dead Code}
As the names suggests, this modification removes code that statically
can be determined to be unreachable. This is standard \llvm\
modification that has to be enabled by the user.

\paragraph{Constant Propagation}
This is a standard \llvm\ modification that forwards constants in the
\llvm-code and thereby reduce the number of instructions in the
\llvm-code.  

\paragraph{Mem2Reg}
This modification tries to promote memory operations to register
operations. This is useful as it makes operations easier for some of
the modifications. The modification can be enabled by the user or
\emph{forced} by other modifications.

\paragraph{Loop Unrolling}
This is the only modification that requires a user specified input
$n$. The modification unrolls all detected loops in the program at most $n$
times. If it can be determined a loop will only execute $m < n$ times,
it is of course only unrolled $m$ times. The unrolling is implemented
inside \lodin\ but borrows the unrolling strategy from the \llvm\
library. The reason the loop unrolling does not use the default \llvm\
unrolling method is that \lodin\ needs more control of the unrolling
than the interface offered. Enabling loop unrolling force-enables \emph{Mem2Reg} and
\emph{Simplify CFG}. The main usage of Loop unrolling is to support the unrolling needed by
bounded model checking.

\subsection{Architecture}
\lodin\ employs a layered architecture (see \autoref{fig:architecture}) where high-level algororithms -
as detailed in previous sections - can be implemented without
knowledge of low-level consideratins such as how the states are
represented. The algorithms depends on state generators implementing
the the state space reductions or the probabilistic semantics. The
generators in turns depends on a joint interpreter-platform unit, that
will interact with an interface to a state representation (how
activation records are stored etc.). The  state
representation then depends on a context-memory unit which performs
the operations requested by the interpreter. At the lowest level of
the architecture is the storage unit which is responsible for storing
and saving states (used by the implementation of \passed/\waiting\ sets
in \autoref{alg:reachFixP}). 
\definecolor{mybluei}{RGB}{124,156,205}
\definecolor{myblueii}{RGB}{73,121,193}
\definecolor{mygreen}{RGB}{202,217,126}
\definecolor{mypink}{RGB}{233,198,235}

\newcommand\widernode[5][widebox]{
  \node[
    #1,
    fit={(#2) (#3)},
    label=center:{\sffamily\bfseries\color{white}#4}] (#5) {};
  }
  
  \begin{figure}[tb]
    \centering
    \begin{tikzpicture}[node distance=3pt,outer sep=0pt,
boxstyle/.style={
  draw=white,
  fill=#1,
  rounded corners,
  font={\sffamily\bfseries\color{white}},
  align=center,
  minimum height=30pt
},
box/.style={
  boxstyle=#1,
  text width=2.5cm},
box/.default=mybluei,
title/.style={font={\sffamily\bfseries\color{white}}},
widebox/.style={draw=white,inner sep=0pt, rounded corners,fill=#1},
widebox/.default=mybluei,
mylabel/.style={font={\sffamily\bfseries\color{white}}},
]

\matrix (stack) [boxstyle=mybluei!40, draw=black,%
        column sep=3pt, row sep=3pt, inner sep=4mm,%
        matrix of nodes,%
        nodes={box, outer sep=0pt, anchor=center, inner sep=3pt},%
        nodes in empty cells,
        row 1/.style={nodes={fill=none,draw=none,minimum height=3mm}},
        ]
        {
          &  \\
          Generators & Prob-Generators \\
          Interpreter & Platforms \\
          &  \\
          Context & Memory \\
          &  \\
        };
\widernode{stack-1-1}{stack-1-2}{Algorithms}{Algo}
\widernode{stack-4-1}{stack-4-2}{State Rep}{StateRep}
\widernode{stack-6-1}{stack-6-2}{Storage}{Store}

\end{tikzpicture}
\caption{Architecture of \lodin}\label{fig:architecture}
\end{figure}
.

\paragraph{SMT Solvers}
\lodin\ uses external SMT-solvers  for solving the contraints gathered
by the symbolis context implementation. The constraints are
represented in a solver-independent format and only at the last minute
converted to SMT-solver specifics. This allows easily interchanging
the used solver: currently \lodin\ is linked against \zthree~\citep{DBLP:conf/tacas/MouraB08} and
\boolector~\citep{DBLP:conf/cav/NiemetzPWB18} and uses \boolector\ by
default.


\section{Conclusion}
We presented the fairly new tool \lodin{}. \lodin\ implements
explicit-state model checking of \llvm with concurrent processes. To combat the state-space
explosion problem \lodin\ supplements explicit-state model checking
techniques  with simulation-based techniques.  For single-threaded programs \lodin\ implements a symboic state space
representation allowing it to verify programs with non-deterministic
input precisely. The symbolic enigne of \lodin\ uses off-the-shelf
SMT-solvers - presently \boolector\ and \zthree.

\bibliographystyle{plainnat}
\bibliography{bibliography}

\end{document}